\documentclass[aps,pre,twocolumn,showpacs,preprintnumbers,amsmath]{revtex4}
\usepackage{graphicx}%
\usepackage{bm}%
\usepackage{mathptm}%
\usepackage{color}%
\usepackage{ifthen}%

\def\jmdbox#1{{#1}}
\long\def\ignore#1{}

\begin{document}
\preprint{APS/V0.1}
\title{Variational Perturbation Theory for Fokker-Planck Equation with Nonlinear Drift}
\author{Jens Dreger$^1$}
\email[]{jens.dreger@physik.fu-berlin.de}
\author{Axel Pelster$^2$}
\email[]{pelster@uni-essen.de}
\author{Bodo Hamprecht$^1$}
\email[]{bodo.hamprecht@physik.fu-berlin.de}
\affiliation{%
$^1$Freie Universit{\"a}t Berlin, Institut f{\"u}r Theoretische Physik,
Arnimallee 14, 14195 Berlin, Germany\\
$^2$Fachbereich Physik, Universit{\"a}t Duisburg-Essen, Universit{\"a}tsstrasse 5, 45117 Essen, Germany}
\date{\today}
\begin{abstract}
We develop a recursive method for perturbative solutions of the 
Fokker-Planck equation with nonlinear drift. The series expansion of the time-dependent 
probability density in terms of powers of the coupling constant is obtained by solving a set of first-order 
linear ordinary differential equations. Resumming the series in the spirit of variational 
perturbation theory we are able to determine the probability density 
for all values of the coupling constant. Comparison with numerical results shows exponential 
convergence with increasing order.
\end{abstract}
\maketitle
\newcommand{\mP}{\mathcal{P}}
\newcommand{\mPs}{\mathcal{P}_{\text{stat}}}
\newcommand{\mNvpt}{\mN_{\text{VPT}}}
\newcommand{\mNs}{\mN_{\text{strong}}}
\newcommand{\mNw}{\mN_{\text{weak}}}
\newcommand{\mN}{\mathbf{N}}
\newcommand{\wopt}{\omega_{\text{opt}}}
\newcommand{\kopt}{\kappa_{\text{opt}}}
\newcommand{\mPot}{\Phi}
\newcommand{\mSub}[1]{{\em #1}.}
\section{\label{sec:1}INTRODUCTION}
In many problems of physics, chemistry and biology one has to deal with
vast numbers of influences which are not fully known and are thus modelled by
noise or fluctuations \cite{Gardiner,Stratonovich,Kampen,Haken1}. 
The stochastic approach to such systems identifies some relevant macroscopic property $x$ governed by a drift coefficient $K(x)$, while
the microscopic degrees are averaged to noise, entering as a stochastic force. In the case of additive noise 
this simply leads to a diffusion constant $D$. Thus we arrive at the Fokker-Planck equation (FP) \cite{Risken} for the density $\mP(x,t)$ of the probability to find the system in a state with property $x$ at time $t$:
\begin{equation}
\label{FPE}
\dfrac{\partial}{\partial t}\mP(x,t)=-\frac{\partial}{\partial x}\left[K(x) \mP(x,t)\right]+ 
D\frac{\partial^2}{\partial x^2}\mP(x,t) \, .
\end{equation}
Once  $\mP(x,t)$ is known from solving (\ref{FPE}), 
we can calculate the ensemble average of any function $\mathcal{O}(x)$ according to
\begin{eqnarray}
\langle\mathcal{O}(x(t))\rangle = \int_{-\infty}^{+\infty}\mathcal{O}(x(t)) \mP(x,t)\, d x \, .
\end{eqnarray}
In this paper we consider a stochastic model with the nonlinear drift coefficient
\begin{eqnarray}
\label{ANHARM}
K(x) = -\gamma\, x - g x^3 \, .
\end{eqnarray}
Such models  are studied, for instance, in semiclassical treatments of a laser near to its instability  threshold \cite{Risken,Haken3},
where the variable $x$ is taken to be the electric
field. The damping constant $\gamma$ is set proportional to the difference between
the pump parameter $\sigma$ and its threshold value $\sigma_{\rm thr}$, so that the laser instability corresponds to $\gamma = 0$.
The coupling constant $g \geq 0$ describes the interaction between light and
matter within the dipole approximation, and the diffusion constant $D$ in the FP equation
characterizes the spontaneous emission of radiation.
While the linear case with $g=0$, corresponding to Brownian motion with  damping constant $\gamma$,
can be solved analytically, there is no exact solution for the nonlinear case with $g > 0$. But we can find a solution in form of a series expansion of the probability density $\mP(x,t)$ in powers of the coupling
constant $g$. This series is asymptotic, i.e.~its expansion coefficients increase factorially
with the perturbative oder and alternate in sign. \\

Such divergent weak-coupling series are known from various fields of
physics, e.g. quantum statistics or critical phenomena, and resummation techniques have been invented to extract meaningful information in such situations. 
Powerful tools among them are the variational methods, independently 
studied by many groups. (see e.g. the references in \cite{HampKleinert1}). A simple example is the so-called $\delta$-expansion of the anharmonic quantum oscillator, where the trial frequency of an artificial  harmonic oscillator introduced to maximally counterbalance the nonlinear term, is optimized following the  
principle of minimal sensitivity \cite{Stevenson}. It turns out that the $\delta$-expansion procedure 
corresponds to a systematic extension of a variational approach
in quantum statistics \cite{Feynman1,Feynman2,Tognetti1,Tognetti2} 
to arbitrary orders as developed by Kleinert
\cite{Kleinertsys,Festschrift,Kleinert}, now being called variational perturbation theory (VPT).
In recent years, VPT has been extended in a simple but essential way to also allow for the 
resummation of divergent perturbation expansions which arise
from renormalizing the $\phi^4$-theory of critical phenomena \cite{Festschrift,KleinertD1,KleinertD2,Verena}.
The most important new feature of this field-theoretic
variational perturbation theory
is that it
accounts for
the anomalous power approach
to the strong-coupling limit which the
$\delta$-expansion
cannot do. This approach is governed by an irrational
critical exponent
as was first shown by Wegner \cite{Wegner}
in the
context of critical
phenomena.
In contrast to the
$\delta$-expansion, the
field-theoretic
variational perturbation
expansions cannot be derived from 
adding and  subtracting a harmonic term. Instead, a self-consistent procedure is set up to determine
this irrational critical Wegner exponent.
The theoretical results of the field-theoretic variational perturbation theory are in excellent 
agreement with the only
experimental value available so far with appropriate accuracy, the
critical exponent $\alpha$ governing the behaviour of the specific heat 
near the superfluid phase transition of
${}^4$He which was measured in a satellite orbiting around the
earth~\cite{Festschrift,Verena,Satellit,Lipa}.\\

Recently, the VPT techniques have
been applied to Markov theory, approximating a nonlinear stochastic process by an
effective Brownian motion \cite{Putz,Okopinska}.
This is achieved by adding and subtracting a linear term to the nonlinear drift coefficient (\ref{ANHARM}),
where the new damping constant is taken as the variational parameter.
In the present paper we extend this result to higher orders, which we have made accessible by our recursive approach \cite{Dreger}. By doing so,
we are able to show that VPT makes Markov theory converge exponentially with respect to order, a phenomenon known for various other systems \cite{Janke1,Janke2,Guida,Proof,Kuerzinger,Weissbach,Schanz}. \\

The paper is structured as follows. In Sec.~\ref{sec:fp} we review some properties of the
FP equation which are essential for our discussion.
In Sec.~\ref{sec:norm} we present the asymptotic perturbation expansion for the normalization constant
of the stationary solution of the FP equation with the drift coefficient (\ref{ANHARM}) 
and its variational resummation. This simple case already illustrates in an introductory fashion all features of the upcoming treatment of the
time-dependent problem. In Sec.~\ref{sec:rec} we perturbatively solve
the FP equation with a nonlinear drift coefficient (\ref{ANHARM})  for Gaussian initial distributions
by means of a double expansion with respect to the coupling 
strength $g$ and the variable $x$. In Sec.~\ref{sec:vpt} we apply VPT to 
the resulting divergent weak-coupling expansion of the probability density to render the results
convergent for all values of the coupling constant $g$. 
Furthermore, we discuss the exponential convergence of our variational method with respect to the order.
The paper closes with a summary in Sec.\ref{sec:concl}.
\section{\label{sec:fp}FOKKER-PLANCK EQUATION}
In this section we fix our notation by
reviewing the main properties of one-dimensional stochastic Markov processes. 
\subsection{Definitions} 
A stochastic Markov process $x ( t )$ is described by an ordinary stochastic differential equation which is of the Langevin form
\begin{equation}
\label{eq:langevin}
\dot{x}(t) =a(x(t))+b(x(t)) \, \Gamma(t) \, ,
\end{equation}
where $\Gamma(t)$ is a Gaussian distributed stochastic force with zero mean and $\delta$-correlation:
\begin{equation}
\langle\Gamma(t)\rangle=0,\qquad \langle\Gamma(t)\Gamma(t')\rangle=2\delta(t-t')\,.
\end{equation}
For the special case of the coefficient $b(x)\equiv b$ being constant, the stochastic system
is said to be driven by additive noise, and the time evolution
of its probability density $\mP(x,t)$ is described by the Fokker-Planck equation (\ref{FPE})
with drift coefficient $K(x)=a(x)$ and diffusion coefficient $D=b^2$ \cite{Risken}.
The FP equation (\ref{FPE}) has the form of a continuity equation 
\begin{eqnarray}
\frac{\partial}{\partial t} \mP(x,t) + \frac{\partial}{\partial x} \mathcal{S}(x,t) = 0
\end{eqnarray}
with probability current 
\begin{eqnarray}
\mathcal{S}(x,t) = K(x) \mP(x,t) - D\frac{\partial}{\partial x}\mP(x,t)\,. 
\end{eqnarray}
For natural boundary conditions, where $\mathcal{S}(x,t) \to 0$ as $x\to\pm\infty$, probability conservation is thus guaranteed:
\begin{equation}
\frac{\partial}{\partial t}\int_{-\infty}^{+\infty} d x\, \mP(x,t) = 0\, .
\end{equation}
In the case of additive noise, the drift coefficient $K(x)$ can be derived from a potential 
\begin{equation}
\label{POTT}
\mPot(x) = -\frac{1}{D}\int^x K(y)\, d y
\end{equation}
such that
\begin{eqnarray}
K(x)=-D\,\mPot'(x) \, .
\end{eqnarray}
If the potential satisfies $\mPot(x)\to+\infty$ for $x\to\pm \infty$,
the probability density $\mP(x,t)$ approaches a stationary values $\mPs(x)$ in the long-time limit, i.e.
\begin{equation}
\label{STATT}
\mPs(x) = \lim_{t \to \infty}\mP(x,t) =\mN\, e^{-\mPot(x)} \, ,
\end{equation}
where the normalization constant $\mN$ is found to be
\begin{equation}
\mN = \left\{\int_{-\infty}^{+\infty} \exp \left[ \frac{1}{D} \int^x K(y)\, d y \right] d x\right\}^{-1}\,.
\end{equation}
\subsection{Brownian Motion} 
For Brownian motion, defined by a linear drift coefficient
\begin{equation}
\label{LD}
K(x) = - \gamma\, x
\end{equation}
with damping $\gamma > 0$,  we have a harmonic potential: 
\begin{equation}
\mPot(x) =  \dfrac{\gamma}{2D}x^2 
\end{equation}
and the FP equation (\ref{FPE}) has a solution in closed form.
With initial condition
\begin{eqnarray}
\mP(x,x_0;0) = \delta(x-x_0)
\end{eqnarray}
the solution is
\begin{eqnarray}
\mP(x,x';t)&=&\sqrt{\frac{\gamma}{2\pi D (1-e^{-2\gamma\, t})}} \nonumber \\
&& \times \exp\!\left[-\frac{(x-x' e^{-\gamma\, t})^2}{2 D (1-e^{-2\gamma\, t})/\gamma}\right] \, .
\end{eqnarray}
It represents the Green function of the FP equation with the drift coefficient (\ref{LD}), describing the evolution of the probability density starting with an arbitrary initial distribution according to
\begin{eqnarray}
\mP(x,t) = \int_{-\infty}^{+ \infty} \mP(x,x';t-t') \mP(x',t') d x' \, .
\end{eqnarray}
For instance, the initial Gaussian probability density
\begin{equation}
\mP (x;t=0)= \frac{1}{\sqrt{2 \pi \sigma^2}} 
\exp\left[-\frac{(x-\mu)^2}{2 \sigma^2}\right]
\label{harm-normal}
\end{equation}
has the time evolution
\begin{eqnarray}
\label{harm-sol1}
\mP_{\gamma}(x,t)&=&\sqrt{\frac{\gamma}{2\pi [D-(D-\gamma \sigma^2) e^{-2\gamma t}]}} \nonumber \\
&&\times \exp\!\left[-\frac{\gamma}{2}\frac{(x-\mu\, e^{-\gamma t})^2}{D-(D-\gamma \sigma^2)e^{-2\gamma t}}\right]\,.
\end{eqnarray}
In the long-time limit it approaches the stationary distribution 
\begin{eqnarray}
\mPs(x) = \lim_{t \to \infty} \mP_{\gamma}(x,t)
= \sqrt{\frac{\gamma}{2 \pi D}} \exp\left(-\frac{\gamma\, x^2}{2 D}\right) 
\end{eqnarray}
which turns out to be independent of the parameters $\mu$ and $\sigma$ of the initial distribution (\ref{harm-normal}).
\subsection{Nonlinear Drift} 
Consider now an additional nonlinear drift (\ref{ANHARM}) so that the potential (\ref{POTT}) becomes
\begin{equation}
\label{NONPO}
\Phi(x) = \frac{\gamma}{2D} x^2 + \frac{g}{4D} x^4 \,.
\end{equation}
While there exists no closed solution of the corresponding FP equation (\ref{FPE})
\begin{equation}
\label{FPNA}
\frac{\partial}{\partial t} \mP(x,t) = 
\frac{\partial}{\partial x} \left[ (\gamma x + g x^3 )  \mP(x, t) \right] +
D \frac{\partial^2}{\partial x^2} \mP( x, t)\, ,
\end{equation}
its stationary distribution (\ref{STATT}) is given by
\begin{equation}
\label{STAT}
\mPs(x) = \mN(g) \exp \left( - \frac{\gamma}{2 D} x^2 - \frac{g}{4 D} x^4 \right) 
\end{equation}
with normalization constant 
\begin{eqnarray}
\label{NOO}
\mN (g) = \left[\int_{-\infty}^{+\infty} \exp \left( - \frac{\gamma}{2 D}x^2 - \frac{g}{4 D} x^4 \right)  d x\right]^{-1} \, .
\end{eqnarray}
Performing the integral in (\ref{NOO}), we obtain
\begin{equation}
\mN(g)=\begin{cases}
\sqrt{\frac{2 g}{\gamma}} \exp\left(-\tfrac{\gamma^2}{8 D g}\right)\Bigl/\mathrm{K}_{1/4}\left( \tfrac{\gamma^2}{8 D g} \right)
& \text{; $\gamma>0$\, ,}\\[.3cm]
\left(\frac{4 g}{D}\right)^{1/4}\Bigl/\Gamma\left(\frac{1}{4}\right)
& \text{; $\gamma=0$\, ,}\\[.15cm]
\frac{\frac{2}{\pi}\sqrt{\frac{g}{|\gamma|}} \exp\left(-\tfrac{\gamma^2}{8 D g}\right)}
{\left[ \mathrm{I}_{- 1/4} \left( \tfrac{\gamma^2}{8 D g} \right) +\mathrm{I}_{1/4}
\left( \tfrac{\gamma^2}{8 D g}\right) \right]}
& \text{; $\gamma<0$\,,}
\end{cases}
\end{equation}
where $\mathrm{I}_{\nu}(z)$ and $\mathrm{K}_{\nu}(z)$ denote modified Bessel functions
of $\nu$-th order of first and second kind, respectively \cite{Gradshteyn}. 
\begin{table}[t!]
\begin{tabular}{|c|c|c|}
\hline
\rule[-3pt]{0pt}{5mm}
$n$ & \lower5pt\hbox to 0pt{\vrule width 0pt height 16pt}\hfil$\sqrt{2\pi}\,a_n$\hfil &\hfil$b_n$\hfil \\[1mm] \hline \hline
\rule[-3pt]{0pt}{5mm}
0 & $1$ &  \hspace*{2mm} $0.390\,062\,251\,089\,407 = \Gamma(3/4)/\pi$  \hspace*{2mm}  \\[1mm] \hline
\rule[-3pt]{0pt}{5mm}
1 & $3/4 $ &$0.131\,836\,797\,004\,050\,253\,244$ \\[1mm] \hline
\rule[-3pt]{0pt}{5mm}
2 & $-87/32$ &$-0.004\,198\,378\,378\,722\,963\,623$ \\[1mm] \hline
\rule[-3pt]{0pt}{5mm}
3 & $2889/128$ &$-0.001\,419\,006\,213\,792\,844\,574$ \\[1mm] \hline
\rule[-3pt]{0pt}{5mm}
4 & $-581157/2048$ &$0.000\,536\,178\,450\,689\,882\,683$ \\[1mm] \hline
\rule[-3pt]{0pt}{5mm}
5 & \hspace*{2mm} $38668509/8192$ \hspace*{2mm}& $-0.000\,093\,437\,511\,028\,762\,876$ \\[1mm] \hline
\end{tabular}
\caption{\label{tab:coeff1}The first 6 coefficients of the weak- and strong-coupling expansion 
(\ref{def:nweak}) and (\ref{def:nstrong}), respectively, for $D=\gamma=1$.} 
\end{table}
\section{\label{sec:norm}NORMALIZATION CONSTANT}
The diverging behaviour of a perturbation series as well as the
method of VPT to overcome
this problem can already be studied by considering the normalization constant (\ref{NOO})
of the stationary solution (\ref{STAT}). To simplify our discussion we assume in this section without loss of generality $D=1$.
\subsection{Weak- and Strong-Coupling Expansion}
\begin{figure}[t!]%
\jmdbox{\includegraphics[width=\columnwidth]{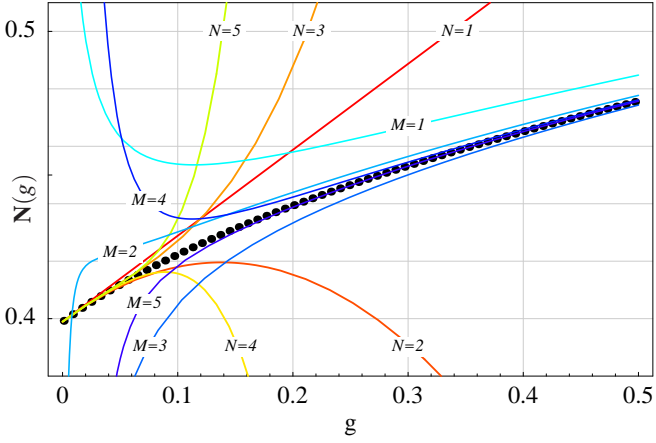}}%
\caption{\label{fig:coeff1}Weak- and 
strong-coupling expansions (\ref{def:nweak}) and (\ref{def:nstrong}) 
up to the $5$th order, respectively, as well as the exact normalization
constant (\ref{NOO}) ($\bullet\,\bullet\,\bullet$) for $D=\gamma=1$.}
\end{figure}

The weak-coupling expansion 
\begin{equation}
\label{def:nweak}
\mNw^{(N)}(g) =  \sum_{n=0}^{N} a_n \, g^n
\end{equation}
follows from (\ref{NOO}) by expanding the $\exp(-g x^4/4)$-term in the integrand:
\begin{equation}
\label{ZWEAK}
\mNw^{(N)}(g) = \left[ \sqrt{2} \sum_{k=0}^N \frac{(-g)^k}{k!} \frac{\Gamma(1/2+2k)}{\gamma^{1/2+2k}} \right]^{-1} \, .
\end{equation}
On the other hand,
a rescaling $x\to (4/g)^{1/4}y$ in the normalization integral (\ref{NOO}) leads to
\begin{equation}
\label{NSUB}
\mN(g) = \left( \frac{g}{4} \right)^{1/4}
\left[\int_{-\infty}^{+\infty} \exp \left( - \frac{\gamma}{g^{1/2}} y^2 - y^4 \right) d y\right]^{-1} \, ,
\end{equation}
so the strong-coupling expansion 
\begin{equation}
\label{def:nstrong}
\mNs^{(M)}(g)= g^{1/4}\sum_{m=0}^{M} b_m \, g^{-m/2} 
\end{equation}
follows from (\ref{NSUB}) by expanding the term $\exp(-\gamma y^2/g^{1/2})$ in the integrand
\begin{equation}
\mNs^{(M)}(g)= \left(\frac{g}{4}\right)^{1/4} \left[\sum_{k=0}^M\frac{(-\gamma)^k}{g^{k/2}} \frac{\Gamma(1/4+k/2)}{2k!}\right]^{-1}.
\end{equation}
Note that the weak-coupling expansion (\ref{def:nweak}) contains integer powers of $g$, whereas dimensional arguments lead to
rational powers of $g$ for the strong-coupling expansion (\ref{def:nstrong}).
The first six coefficients $a_n, b_m$ are given in Table \ref{tab:coeff1}
and the respective expansions (\ref{def:nweak}) and (\ref{def:nstrong}) are 
depicted in Figure \ref{fig:coeff1}. While the weak-coupling expansion provides good
results for small values of $g$, the strong-coupling expansion describes the behavior for large values of $g$.
For intermediate values of $g$ both series yield poor results. \\
\subsection{Variational Perturbation Theory}\label{VPTNORM}
Despite its diverging nature, all information on the analytic function $\mN(g)$ is already contained
in the weak-coupling expansion (\ref{def:nweak}). One way to extract this information and use it to render
the series convergent for any value of the coupling constant $g$ 
is provided by VPT as developed by Kleinert \cite{Verena,Festschrift,Kleinert}. 
This method is based on introducing a dummy variational parameter
on which the full perturbation expansion does
not depend, while the truncated expansion does. 
The optimal variational parameter is then selected by invoking the principle
of minimal sensitivity \cite{Stevenson},  requiring the
quantity of interest to be stationary with respect to the
variational parameter.
In our context \cite{Putz,Okopinska,Dreger}, this dummy variational parameter can be thought of as
the damping constant $\kappa$ of a trial Brownian motion with a harmonic potential $\kappa x^2/2$,
which is tuned in such a way, that it effectively compensates the nonlinear potential.
In order to introduce the variational parameter $\kappa$,
we add the harmonic potential $\kappa x^2/2$ of the trial Brownian motion to the nonlinear potential (\ref{NONPO}) and
subtract it again:
\begin{equation}
\label{NONPOB}
\Phi(x)=\frac{\kappa}{2} x^2 + \frac{g}{4} x^4 + \frac{\gamma -\kappa}{2} x^2 \, .
\end{equation}
By doing so, we consider the harmonic potential $\kappa x^2/2$ as the unperturbed term and treat all remaining potential
terms in (\ref{NONPOB}) as a perturbation. Such a formal perturbation expansion is performed for the normalization
constant (\ref{NOO}) according to
\begin{eqnarray}
\mN(g)&=& \left\{ \int_{-\infty}^{+\infty} 
\exp \Bigg[ -\frac{\kappa}{2} x^2 
\right. \nonumber \\ && \left. \left. 
- \delta \left(\frac{g}{4} x^4 + \frac{\gamma - \kappa}{2} x^2\right) \Bigg]
d x \right\}^{-1} \right|_{\delta=1} \, ,
\end{eqnarray}
where the additional parameter $\delta$ is introduced in the spirit of the $\delta$-expansion 
(see, for instance, the references in \cite{HampKleinert1}).
Due to the rescaling $x\to \tilde{x}/\beta(\delta)$ with the scaling factor 
\begin{equation}
\label{BFAC}
\beta(\delta)=\sqrt{[\kappa + \delta( \gamma - \kappa)]/\gamma}\,,
\end{equation}
the weak-coupling expansion (\ref{def:nweak}) of (\ref{NOO}) leads to:
\begin{equation}
\mN(g)=\left.  \beta ( \delta) 
\sum_{n=0}^{\infty} a_{n} \left[ \frac{\delta g}{\beta^4 (\delta)} \right]^{n}  \right|_{\delta=1} \, . 
\end{equation}
Expanding and truncating the series at order $N$ in $\delta$, we
obtain the $N$th variational approximation to the normalization
constant $\mN(g)$:
\begin{eqnarray}
\mNvpt^{(N)}(g,\kappa) &=& 
\sum_{n=0}^N a_n \left(\frac{\kappa}{\gamma}\right)^{1/2-2n} g^n \nonumber \\
&& \times \sum_{k=0}^{N-n}
\left( \begin{array}{c}  1/2-2n\\ k \end{array} \right) 
\left(\frac{\gamma-\kappa}{\kappa}\right)^{k} \, .
\label{TRUNC}
\end{eqnarray}
Equivalently, the variational expression (\ref{TRUNC}) also follows directly from the weak-coupling expansion (\ref{def:nweak}).
To this end we remark that, due to dimensional arguments, the respective coefficients depend via 
\begin{eqnarray}
\label{AN}
a_n = \tilde{a}_n \, \gamma^{1/2-2n}
\end{eqnarray}
on $\gamma$ where $\tilde{a}_n$ denote dimensionless quantities. Treating in (\ref{NONPOB}) the harmonic potential $\kappa x^2/2$
as the unperturbed term and all remaining potential terms as a perturbation, corresponds then to the substitution
\begin{eqnarray}
\label{SUB}
\gamma \to \kappa ( 1 + g r )
\end{eqnarray}
with the abbreviation
\begin{eqnarray}
\label{ABB}
r = \frac{\gamma-\kappa}{\kappa g} \, .
\end{eqnarray}
Thus inserting (\ref{SUB}) in the weak-coupling expansion (\ref{def:nweak}), (\ref{AN}), a reexpansion in powers of $g$ up to order
$N$ together with the resubstitution (\ref{ABB}) leads to a rederivation of (\ref{TRUNC}).\\ 

As the normalization constant $\mN(g)$ does not depend on the variational parameter $\kappa$, it is reasonable to ask for the truncated
series (\ref{TRUNC}) to depend on $\kappa$ as little as possible. Therefore, we invoke the principle of minimal sensitivity
\cite{Stevenson} and demand
\begin{equation}
\label{FIRST}
\left.\frac{\partial}{\partial \kappa} \mNvpt^{(N)}(g,\kappa)\right|_{\kappa=\kopt^{(N)}(g)} = 0\,.
\end{equation}
The optimal variational parameter $\kopt^{(N)}(g)$ leads to the $N$th order variational approximation 
$\mNvpt^{(N)}(g,\kopt^{(N)}(g))$ to the normalization constant  $\mN(g)$. In case that (\ref{FIRST}) is not solvable, we determine
$\kopt^{(N)}(g)$ from the zero of the second derivative in accordance with 
the principle of minimal sensitivity
\cite{Stevenson}:
\begin{equation}
\label{SECOND}
\left.\frac{\partial^2}{\partial \kappa^2} \mNvpt^{(N)}(g,\kappa)\right|_{\kappa=\kopt^{(N)}(g)} = 0\,.
\end{equation}
If it happens that (\ref{FIRST}) or (\ref{SECOND}) have more than one solution, we select that particular one which is
closest to the solution in the previous variational order. \\

\begin{figure}[t!]
\jmdbox{\includegraphics[width=\columnwidth]{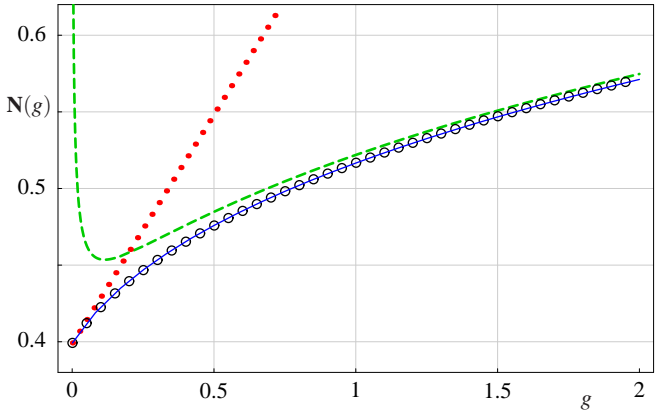}}%
\caption{\label{fig:vptnorm1}First-oder variational result ({\bf{\color{blue}---}}) compared with exact 
normalization constant $(\circ\circ\circ)$. First-order weak- and strong-coupling approximations are shown by 
dashed ({\bf\color[rgb]{0,.8,0}--\,--}) and dotted ({\color[rgb]{.8,0,0}$\cdot\!\cdot\!\cdot$}) lines, respectively.}%
\end{figure}
As an illustrative example we treat explicitly the first variational order where we obtain
\begin{equation} 
\label{vptord1-Nw1} 
\mNvpt^{(1)}(g,\kappa) =
\frac{2 \kappa (\kappa + \gamma) + 3 g}{4 \sqrt{2 \pi \kappa^{3}}}\,,
\end{equation}
so that the first derivative with respect to $\kappa$
\begin{equation}\label{vptord1-Nw1d1}
\frac{\partial}{\partial \kappa}\mN^{(1)}_{\text{vpt}}(g,\kappa) = 
\frac{2\kappa (\kappa - \gamma) - 9 g}{8\sqrt{2\pi \kappa^5}}
\end{equation}
has the two zeros
\begin{equation}
\label{vptord1-woptsol}
\kopt^{(1,\pm)}(g) = \frac{1}{2}\left(\gamma \pm \sqrt{18 g + \gamma^2}\,\right).
\end{equation}
Since the variational parameter has to approach $\gamma$ for a vanishing coupling constant $g$,
we select from (\ref{vptord1-woptsol}) the solution $\kopt^{(1,+)}(g)$. 
The resulting optimized result $\mNvpt^{(1)}(g,\kopt^{(1,+)}(g))$
is shown in Figure \ref{fig:vptnorm1}. We observe that in this parameter range already the first-order variational approximation 
$\mNvpt^{(1)}(g,\kopt^{(1,+)}(g))$
is indistinguishable from the exact normalization constant $\mN(g)$.
\subsection{Exponential Convergence}
In order to quantify the accuracy of the variational approximations, we study now, in particular, the strong-coupling regime
$g \to \infty$. In first order, the insertion of (\ref{vptord1-woptsol}) in (\ref{vptord1-Nw1}) 
leads to the strong-coupling expansion (\ref{def:nstrong}) with the leading coefficient
\begin{eqnarray}
\label{VAL}
b_0^{(1)} = \left( \frac{2}{9 \pi^2} \right)^{1/4} \approx 0.387 \, .
\end{eqnarray}
Comparing (\ref{VAL}) with $b_0=\Gamma(3/4)/\pi \approx 0.390$ 
(see Table \ref{tab:coeff1}), we conclude that first-order variational perturbation theory
yields the leading strong-coupling coefficient within an accuracy of less than 1 \%.\\

\begin{figure}[t!]
\jmdbox{\includegraphics[width=\columnwidth]{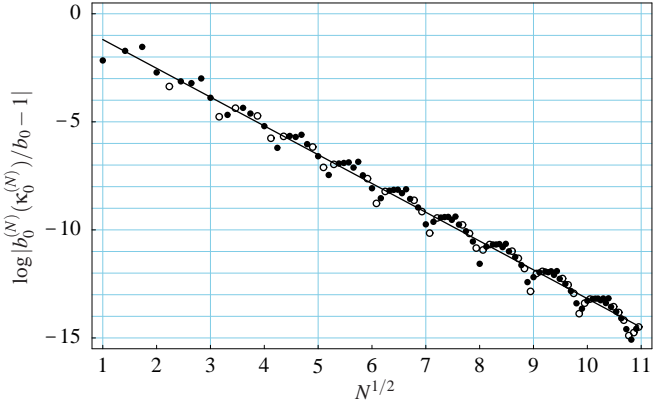}}%
\caption{\label{log}The points show the logarithmic plot of the relative error $|b^{(N)}_0  (\kappa_0^{(N)})- b_0|/b_0$
when using the smallest zero of the first ($\bullet$) and the second derivative $(\circ)$, respectively,
against $N^{1/2}$. The solid line represents a fit to the straight line $- \alpha - \beta N^{1/2}$.}
\end{figure}

To obtain higher-order variational results for this strong-coupling coefficient $b_0$, we proceed as follows.
From the first-order approximation  (\ref{vptord1-woptsol}) we see that the variational parameter has a strong-coupling
expansion $\kappa^{(1,+)}_{\rm opt} ( g ) = \sqrt{g} \, 3 \sqrt{2}/2 + \ldots$,
whose form turns out to be valid also for the orders $N>1$.
Inserting the ansatz $\kappa^{(N)}_{\rm opt} ( g ) = \kappa_0^{(N)} \sqrt{g} + \ldots$
in (\ref{TRUNC}), we obtain the $N$th order approximation for the leading strong-coupling
coefficient $b_0$:
\begin{eqnarray}
b_0^{(N)} (\kappa_0^{(N)}) &=& \sum_{n=0}^N 
a_n \left( \frac{\kappa_0^{(N)}}{\gamma} \right)^{1/2-2n} \nonumber \\
&&\times \sum_{k=0}^{N-n} \left( \begin{array}{@{}c} 1/2 - 2 n \\ k \end{array} \right)
\, (-1)^k \, .
\end{eqnarray}
The inner sum can be performed explicitly by using Eq. (0.151.4) in Ref. \cite{Gradshteyn}:
\begin{eqnarray}
\label{TTT}
\hspace*{-0.2cm} b_0^{(N)} (\kappa_0^{(N)}) =\sum_{n=0}^N  (-1)^{N-n} \, \left( \begin{array}{@{}c} - 1/2 - 2n 
\\ N-n \end{array} \right) a_n \left( \frac{\kappa_0^{(N)}}{\gamma} \right)^{1/2-2n} \hspace*{-0.6cm} .
\end{eqnarray}
In order to optimize (\ref{TTT}) we look again for an extremum
\begin{eqnarray}
\label{1}
\frac{\partial b_0^{(N)} (\kappa_0^{(N)})}{\partial \kappa_0^{(N)}} = 0 
\end{eqnarray}
or for a saddle point
\begin{eqnarray}
\label{2}
\frac{\partial^2 b_0^{(N)} (\kappa_0^{(N)})}{\partial \kappa_0^{(N)} {}^2} = 0 \, .
\end{eqnarray}
It turns out that extrema exist for odd orders $N$, whereas even orders $N$ lead to saddle points.
The points of Figure \ref{log}
show the logarithmic plot of the relative error $|b^{(N)}_0 (\kappa_0^{(N)}) - b_0|/b_0$ when using the smallest zero
of the first and second derivative, i.e.~(\ref{1}) and (\ref{2}), respectively.
We observe that the relative error depends linearily on $N^{1/2}$ up to the order $N=120$ according to
\begin{eqnarray}
\label{EXP}
\frac{|b^{(N)}_0 (\kappa_0^{(N)}) - b_0|}{b_0} = e^{- \alpha - \beta \, N^{1/2} }\, ,
\end{eqnarray}
where the fit to the straight line $- \alpha - \beta N^{1/2}$
leads to the quantities $\alpha = 0.139714$ and $\beta = 1.33218$. Thus we have demonstrated that the
variational approximations for the leading strong-coupling coefficient converge exponentially fast.
Note that the speed of convergence is faster than the exponential convergence of the variational results 
for the ground-state energy of the anharmonic oscillator \cite{Janke1,Janke2}.
\section{\label{sec:rec}RECURSION RELATIONS}
Now we elaborate the perturbative solution of the FP equation (\ref{FPNA}) 
with the initial distribution (\ref{harm-normal}). By doing so, we follow the notion of Ref.~\cite{Weissbach}
and generalize the recursive Bender-Wu solution method for the 
Schr{\"o}dinger equation of the anharmonic oscillator \cite{Bender}, thus obtaining a recursive set of first-order ordinary 
differential equations \cite{Dreger}.
\subsection{Time Transformation}
At first we perform a suitable time transformation which simplifies the following calculations:
\begin{equation}
\label{trans-def}
\tau(t)=\tau_0 e^{-\gamma t}\, , \qquad \tau_0=\sqrt{1-\gamma \sigma^2 /D}\,.
\end{equation}
Thus the new time $\tau$ runs from $\tau_0$ to $0$ when the physical time $t$ evolves from $0$ to $\infty$. 
Due to (\ref{trans-def}) the FP equation (\ref{FPNA}) is transformed to
\begin{equation}
\label{FPN}
\hspace*{-0.1cm}- \gamma \tau \frac{\partial}{\partial \tau} \mP(x,\tau) = 
\frac{\partial}{\partial x} \left[ (\gamma x + g x^3 )  \mP(x, \tau) \right] +
D \frac{\partial^2}{\partial x^2} \mP( x, \tau)\, .
\end{equation}
Furthermore, the initial distribution (\ref{harm-normal}) reads then
\begin{equation}
\label{coeff-p0t0}
\mP (x,\tau=\tau_0) = \frac{1}{\sqrt{2\pi \sigma^2}}\exp\!\left[-\frac{(x-\mu)^2}{2\sigma^2}\right] 
\end{equation}
and contains $P(x,\tau=1)=\delta(x-\mu)$ as a special case in the limit $\sigma^2\to0$.  
\subsection{Expansion in Powers of $g$}
If the coupling constant $g$ vanishes, the solution of the initial value problem (\ref{FPN}), (\ref{coeff-p0t0})
follows from applying the time transformation (\ref{trans-def}) to (\ref{harm-sol1}), i.e.  
\begin{equation}
\label{trans-sol}
\mP_{\gamma}(x, \tau)=
\sqrt{\frac{\gamma}{2\pi D(1-\tau^2)}}\exp\!\left[-\frac{\gamma}{2 D} \frac{(x-x_0 \tau)^2}{1-\tau^2}\right]\,,
\end{equation}
where we have introduced the abbreviation
\begin{equation}
\label{trans-x0def}
x_0=\frac{\mu}{\sqrt{1-\gamma \sigma^2/D}}\, .
\end{equation}
For a coupling constant $g>0$, we solve (\ref{FPN}) by the ansatz
\begin{equation}
\label{flat-ansatz}
\mP(x,\tau) = \mP_{\gamma} (x,\tau)\,q(x,\tau)\,,
\end{equation}
so that the remainder $q(x,\tau)$ fulfills the partial differential equation
\begin{eqnarray}
&& \hspace*{-1.0cm}-\gamma \tau \frac{\partial}{\partial \tau} q(x,\tau)=
\left[3 g x^2 + \frac{\gamma gx^4-\gamma g \tau x_0 x^3}{D(\tau^2-1)} \right] q(x,\tau) \nonumber \\
&& \hspace*{-1.0cm}+\left[ \frac{\gamma(\tau^2+1)x-2 \gamma \tau x_0}{\tau^2-1} + g x^3 \right] 
\frac{\partial}{\partial x}q(x,\tau) +
D \frac{\partial^2}{\partial x^2} q(x,\tau)\,.
\label{DIFF}
\end{eqnarray}
Then we solve (\ref{DIFF}) by expanding $q(x,\tau)$ in a Taylor series with respect to the coupling constant $g$, i.e.
\begin{equation}
\label{q-def}
q(x,\tau)=\sum_{n=0}^\infty g^n\, q_n(x,\tau)\,,
\end{equation}
where we set $q_0 (x,\tau)=1$. Thus the expansion coefficients $q_n(x,\tau)$ obey the partial differential equations
\begin{eqnarray}
&& \hspace*{-0.4cm} -\gamma \, \tau \frac{\partial}{\partial \tau}  q_n(x,\tau) = \left[ 3 x^2 
+ \frac{\gamma x^4-\gamma \, \tau x_0 x^3}{D(\tau^2-1)} + \frac{\partial}{\partial x} \right] q_{n-1}(x,\tau)  \nonumber \\
&& \hspace*{-0.4cm}+ \frac{\gamma(\tau^2+2)x-2\gamma \,\tau x_0}{\tau^2-1}  \frac{\partial}{\partial x}  q_n(x,\tau)
+ D \frac{\partial^2}{\partial x^2} q_n(x,\tau)\,. 
\label{DIFFN}
\end{eqnarray}
\subsection{Expansion in Powers of $x$}
It turns out that the partial differential equations (\ref{DIFFN}) are solved by expansion coefficients $q_n(x,\tau)$ 
which are finite polynomials in $x$:
\begin{equation}
\label{phitry}
q_n(x,\tau)=\sum_{k=0}^{M_n} \alpha_{n,k}(\tau) x^k\,.
\end{equation}
Indeed, inserting the decomposition (\ref{phitry}) in (\ref{DIFFN}), we deduce that the highest polynomial degree is 
given by $M_n = 4 n$. 
Note that in case of $\mu=0$ the partial differential equations (\ref{DIFFN}) are symmetric with respect to $x\to-x$,
so that only even powers $x$ appear in (\ref{phitry}). From (\ref{DIFFN}) and
(\ref{phitry}) follows that the functions $\alpha_{n,k}(\tau)$ are determined from
\begin{equation}
\label{coeff-lindgl}
\frac{\partial}{\partial \tau} \alpha_{n,k}(\tau) \,+\, \frac{k(\tau^2+1)}{\tau(\tau^2-1)} \alpha_{n,k}(\tau) = r_{n,k}(\tau)\, ,
\end{equation}
where the inhomogeneity is given by
\begin{eqnarray}
&&\hspace*{-0.6cm} r_{n,k} (\tau)=
-\frac{k+1}{\gamma \,\tau}\alpha_{n-1,k-2}(\tau)-\frac{D(k+2)(k+1)}{\gamma \, \tau}\alpha_{n,k+2}(\tau)
\nonumber \\ && 
\hspace*{-0.6cm}+\frac{x_0\alpha_{n-1,k-3}(\tau)}{D(\tau^2-1)}  
- \frac{\alpha_{n-1,k-4}(\tau)}{D(\tau^3-\tau)} 
+ \frac{2 x_0(k+1)}{\tau^2-1}\alpha_{n,k+1} (\tau)
\,.
\label{coeff-inhomogen}
\end{eqnarray}
\begin{figure}[t]
\jmdbox{\includegraphics[width=\columnwidth]{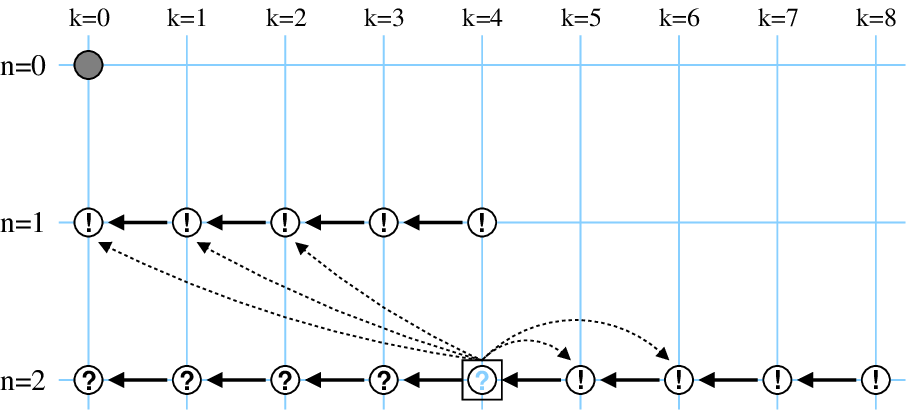}}%
\caption{\label{rekfig}Recursive calculation of the functions $\alpha_{n,k}(\tau)$: $\alpha_{0,0}(\tau)=1$ is the only coefficient
which is a priori nonzero (\protect\raisebox{-.05cm}{\protect\includegraphics[width=2ex]{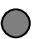}}\!). For each $n$ the coefficients are successively determined for $k=4n,\ldots,0$ (\protect\raisebox{-.06cm}{\protect\includegraphics[width=2ex]{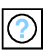}}). Each step necessitates 
(\!\protect\raisebox{.04cm}{\protect\includegraphics[width=4ex]{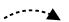}}\!) only those coefficients which are already known (\protect\raisebox{-.04cm}{\protect\includegraphics[width=2ex]{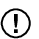}}\!).}
\end{figure}
\subsection{Recursive Solution}
The Eqs. (\ref{coeff-lindgl}) and (\ref{coeff-inhomogen}) represent a system of first-order ordinary differential equations
which can be recursively solved according to Figure \ref{rekfig}. By doing so, one has to take into account 
$\alpha_{n,k}(\tau)=0$ if $n<0$ or $k<0$ or $k>4n$. The respective positions in the $(n,k)$-grid of Figure \ref{rekfig} are empty or
lie outside. Iteratively calculating the coefficients $\alpha_{n,k}(\tau)$ in the $n$th order, we have to start with $k=4n$
and decrease $k$ up to $k=0$. In each iterative step the inhomogeneity (\ref{coeff-inhomogen}) of (\ref{coeff-lindgl}) contains only
those coefficients which are already known. The only coefficient which is a priori nonzero is $\alpha_{0,0}(\tau)=1$.\\

Applying the method of varying constants,
the inhomogeneous differential equation (\ref{coeff-lindgl}) is solved by the ansatz
\begin{eqnarray}
\label{VC}
\alpha_{n,k} (\tau)= A_{n,k}(\tau)\, \frac{\tau^k}{(\tau^2-1)^k} \, ,
\end{eqnarray}
where $A_{n,k}(\tau)$ turns out to be
\begin{equation}
A_{n,k}(\tau)= A_{n,k}(\tau_0) + \int_{\tau_0}^\tau r_{n,k}(\sigma)\,\sigma^{-k}(\sigma^2-1)^{k}\, d \sigma\,.
\end{equation}
The integration constant $A_{n,k}(\tau_0)$ is fixed by considering the initial distribution (\ref{coeff-p0t0}).
From (\ref{trans-sol}), (\ref{flat-ansatz}), (\ref{q-def}), and (\ref{phitry}) follows 
\begin{equation}
\label{coeff-q=1}
1=1+\sum_{n=1}^{\infty} g^n \sum_{k=0}^{4n} \alpha_{n,k}(\tau_0) x^k\, ,
\end{equation}
so we conclude $\alpha_{n,k}(\tau_0)=0$ and thus $A_{n,k} (\tau_0)=0$ for $n \geq 1$. Therefore, we obtain the final result
\begin{equation}
\label{coeff-result-ank}
\alpha_{n,k}(\tau)=\frac{\tau^k}{(\tau^2-1)^k} \int_{\tau_0}^\tau r_{n,k}(\sigma)\,\sigma^{-k}(\sigma^2-1)^{k}\, d \sigma\,,
\end{equation}
where $r_{n,k}(\sigma)$ is given by (\ref{coeff-inhomogen}).\\

Note that the special case $\sigma^2\to0$ with the initial distribution   $P(x,\tau=1)=\delta(x-\mu)$ has to be discussed separately,
as then (\ref{coeff-q=1}) is not valid. 
In this case we still conclude for $k\neq 0$ that $A_{n,k} (\tau_0=1)=0$, otherwise 
$\alpha_{n,k} (\tau)$ in (\ref{VC}) would posses for $\tau=1$ a pole of order $k$. For $k=0$ this argument is not valid
as the fraction in (\ref{VC}) is then no longer present. 
The integration constant $A_{n,0} (\tau_0=1)$ follows from 
considering the normalization integral of (\ref{flat-ansatz}) for $\tau=1$ together with (\ref{q-def}), and (\ref{phitry}), i.e.
\begin{equation}
\label{coeff-q=1B}
1=1+\sum_{n=1}^{\infty} g^n \sum_{k=0}^{4n} \alpha_{n,k}(\tau = 1) x_0^k\, .
\end{equation}
Indeed, taking into account (\ref{VC}) for $k=0$ leads to 
\begin{equation}
A_{n,0}(\tau=1) = \alpha_{n,0}(\tau=1) = -\sum_{k=1}^{4n}\alpha_{n,k}(\tau=1) x_0^k\,.
\end{equation}
Thus for $k \neq 0$ the coefficients $\alpha_{n,k}(\tau)$ are still given by (\ref{coeff-result-ank}), whereas for $k=0$
we obtain 
\begin{equation}
\alpha_{n,0}(\tau) = \int_1^\tau\! r_{n,0}(\sigma)\, d \sigma  -\sum_{k=1}^{4n}\alpha_{n,k}(\tau=1) x_0^k\,.
\end{equation}
\subsection{Cumulant Expansion}\label{CE}
\begin{figure}[t!]
\jmdbox{\includegraphics[width=\columnwidth]{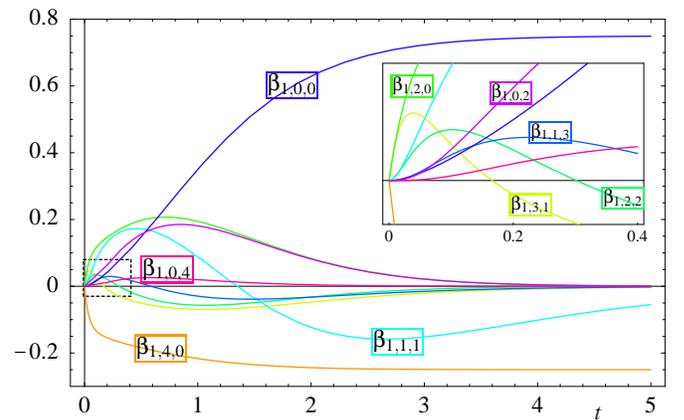}}%
\caption{\label{Result}Time evolution of the expansion coefficients $\beta_{1,k,l}(t)$ from Table II
for $D=1$, $\gamma=1$, $\mu=0$ and
$\sigma=0.5$.}
\end{figure}

The weak-coupling expansion (\ref{flat-ansatz}), (\ref{q-def}) of the probability density $\mP(x,\tau)$
has the disadvantage that its truncation to a certain order $N$ could lead to negative values. To avoid this, we rewrite
the weak-coupling expansion (\ref{flat-ansatz}), (\ref{q-def}) in form of the cumulant 
\begin{equation}
\label{kumulant-ansatz}
\mP(x,\tau) = e^{p(x,t)}\, ,
\end{equation}
where the exponent $p(x,\tau)$ is expanded in powers of the coupling constant $g$:
\begin{equation}
\label{p-def}
p(x,\tau)=\ln \mP_{\gamma}(x,\tau) + \sum_{n=1}^\infty g^n\, p_n(x,\tau)\,.
\end{equation}
The respective coefficients $p_n(x,\tau)$ follow from
reexpanding the weak-coupling expansion (\ref{flat-ansatz}), (\ref{q-def})
according to (\ref{kumulant-ansatz}), (\ref{p-def}). However,
it is also possible to derive a recursive set of ordinary differential equations whose solution directly leads to the
cumulant expansion (\ref{kumulant-ansatz}), (\ref{p-def}). To this end we proceed in a similar way as in case of the 
derivation of the weak-coupling expansion and perform the ansatz
\begin{equation}
\label{coeff-chiansatz2}
p_n(x,\tau)=\sum_{k=0}^{2n+2}\sum_{l=0}^{2n+2-k} \beta_{n,k,l}(\tau) x^k x_0^l\,.
\end{equation}
The respective expansion coefficients $\beta_{n,k,l}(\tau)$ follow from a similar formula than (\ref{coeff-result-ank}), i.e.
\begin{equation}
\label{BETA}
\beta_{n,k,l}(\tau)=\frac{\tau^k}{(\tau^2-1)^k} \int_{\tau_0}^\tau s_{n,k,l}(\sigma)\,\sigma^{-k}(\sigma^2-1)^{k}\, d \sigma \, ,
\end{equation}
where the functions $s_{n,k,l}(\sigma)$ are given for $n=1$ by
\begin{eqnarray}
&& \hspace*{-0.8cm} s_{1,k,l}(\tau)= 
-\frac{3}{\gamma \tau}\delta_{k,2}\delta_{l,0} + 
\frac{\delta_{k,3}\delta_{l,1}}{D(\tau^2-1)}
- \frac{\delta_{k,4}\delta_{l,0}}{D(\tau^3-\tau)}
\nonumber \\
&&\hspace*{-0.8cm} +\frac{2 (k+1)}{\tau^2-1} \beta_{1,k+1,l-1} (\tau)- \frac{D(k+2)(k+1)}{\tau \gamma}\beta_{1,k+2,l}(\tau)\,,
\label{coeff-beta-inhomog-1kl}
\end{eqnarray}
and for $n \geq 2$ by
\begin{eqnarray}
&&s_{n,k,l}(\tau)= 
- \frac{k-2}{\gamma \tau} \beta_{n-1,k-2,l} (\tau)+ \frac{2 (k+1)}{\tau^2-1} \beta_{n,k+1,l-1} (\tau)
\nonumber \\ &&
- \frac{D (k+2)(k+1)}{\gamma \tau}\beta_{n,k+2,l} (\tau)
-\frac{D}{\gamma \tau} \sum_{m=1}^{n-1}\sum_{j=1}^{k+1} j(k-j+2) 
\nonumber \\ &&
\times \sum_{i=0}^{l} \beta_{m,j,i} (\tau) \beta_{n-m,k-j+2,l-i}(\tau)\,.
\label{coeff-beta-inhomog-nkl}
\end{eqnarray}
Iterating (\ref{BETA})--(\ref{coeff-beta-inhomog-nkl}) one has to take into account that $\beta_{n,k,l}(\tau)$
vanishes if one of the following conditions is fulfilled:
$n \leq 0$; $k<0$ or $k> 2n+2$; $l<0$ or $l> 2n+2-k$; $k+l$ odd. By inverting the time transformation (\ref{trans-def}), the
expansion coefficients are finally determined as functions of the physical time $t$.
For the first order $n=1$ one finds the following
expansion coefficients $\beta_{1,k,l}(t)$:
\newcommand{\mystyle}{\normalsize}
\begin{widetext}
{\renewcommand{\arraystretch}{2.3}
\begin{center}
\begin{table}[h!]
\begin{tabular}{ll}
$\beta_{1,0,0}(t)$&$=\mystyle\frac{3 D ( D^2 ( -5 + e^{4 t \gamma } - 4 t \gamma  + 
         e^{2 t \gamma } ( 4 - 8 t \gamma  )  )  + 
      8 D \gamma  ( 1 + t \gamma  + 
         e^{2 t \gamma } ( -1 + t \gamma  )  )  
       {\sigma }^2 - 2 {\gamma }^2 
       ( 1 - e^{2 t \gamma } + 2 t \gamma  )  {\sigma }^4 ) }{4 {\gamma }^2 {( D 
         ( -1 + e^{2 t \gamma } )  + \gamma  {\sigma }^2 ) }^2}$\\
$\beta_{1,0,2}(t)$&$=\mystyle\frac{3 D^2 ( D ( 1 + 
         e^{4 t \gamma } ( -5 + 4 t \gamma  )  + 
         e^{2 t \gamma } ( 4 + 8 t \gamma  )  )  - 
      2 \gamma  ( 1 - e^{4 t \gamma } + 
         4 e^{2 t \gamma } t \gamma  )  {\sigma }^2 ) }{2 \gamma  {( D ( -1 + 
e^{2 t \gamma } )  +         \gamma  {\sigma }^2 ) }^3}$\\
$\beta_{1,0,4}(t)$&$=\mystyle\frac{D^3 ( 1 - 6 e^{2 t \gamma } + 2 e^{6 t \gamma} + 
      e^{4 t \gamma} ( 3 - 12 t \gamma  )  ) }{4 
    {( D ( -1 + e^{2 t \gamma } )  + 
        \gamma  {\sigma }^2 ) }^4}$\\
$\beta_{1,1,1}(t)$&$=\mystyle\frac{-3 D e^{t \gamma } ( 2 D^2 
       ( 4 + 3 t \gamma  + 
         e^{4 t \gamma } ( -2 + t \gamma  )  + 
         e^{2 t \gamma } ( -2 + 8 t \gamma )  )  - 
      D \gamma  ( 13 - e^{4 t \gamma } + 12 t \gamma  + 
         4 e^{2 t \gamma } ( -3 + 4 t \gamma  )  )  {\sigma }^2 + 3 {\gamma }^2 
       ( 1 - e^{2 t \gamma } + 2 t \gamma  )  {\sigma }^4 ) }{2 \gamma  {( D ( -1 
+ e^{2 t \gamma } )  +         \gamma {\sigma }^2 ) }^3}$\\
$\beta_{1,1,3}(t)$&$=\mystyle\frac{-( D^2 e^{t \gamma } 
      ( D ( -1 + e^{6 t \gamma } + 
           e^{4 t \gamma } ( 9 - 12 t \gamma  )  - 
           3 e^{2 t \gamma } ( 3 + 4 t \gamma  )  )  + 3 \gamma  ( 1 - e^{4 t \gamma } + 
           4 e^{2 t \gamma } t \gamma  )  {\sigma }^2 )  ) }{2 {( D ( -1 + e^{2 t \gamma } )  + 
        \gamma  {\sigma }^2 ) }^4}$\\
$\beta_{1,2,0}(t)$&$=\mystyle\frac{3 ( D^3 ( 1 + 
         e^{4 t \gamma } ( -5 + 4 t \gamma )  + 
         e^{2 t \gamma } ( 4 + 8 t \gamma  )  )  - 
      D^2 \gamma  ( 3 + 
         e^{4 t \gamma } ( -7 + 4 t \gamma  )  + 
         4 e^{2 t \gamma } ( 1 + 4 t \gamma  )  )  
       {\sigma }^2 + D {\gamma }^2 
       ( 3 - e^{4 t \gamma } + 
         e^{2 t \gamma } ( -2 + 8 t \gamma  )  )  
       {\sigma }^4 + ( -1 + e^{2 t \gamma } )  {\gamma}^3 
       {\sigma }^6 ) }{2 \gamma  
    {( D ( -1 + e^{2 t \gamma } )  + 
       \gamma  {\sigma }^2 ) }^3}$\\
$\beta_{1,2,2}(t)$&$=\mystyle\frac{-3 D e^{2 t \gamma } 
    ( D^2 ( 3 + 2 t \gamma  + 
         8 e^{2 t \gamma } t \gamma  + 
         e^{4 t \gamma } ( -3 + 2 t \gamma  )  )  - 
      D \gamma  ( 5 - e^{4 t \gamma} + 4 t \gamma  + 
         e^{2 t \gamma } ( -4 + 8 t \gamma )  )  
       {\sigma }^2 + {\gamma }^2 
       ( 1 - e^{2 t \gamma } + 2 t \gamma )  {\sigma }^4 ) }{2 {( D ( -1 + e^{2 t \gamma } )  + 
        \gamma {\sigma }^2 ) }^4}$\\
$\beta_{1,3,1}(t)$&$=\mystyle\frac{e^{t \gamma } ( D^3 
       ( 1 - e^{6 t \gamma } + 
         3 e^{4 t \gamma } ( -3 + 4 t \gamma  )  + 
         3 e^{2 t \gamma} ( 3 + 4 t \gamma  )  )  - 
      3 D^2 \gamma  ( 1 + 
         e^{4 t \gamma } ( -5 + 4 t \gamma  )  + 
         e^{2 t \gamma } ( 4 + 8 t \gamma  )  )  
       {\sigma }^2 + 3 D {\gamma }^2 
       ( 1 - e^{4 t \gamma} + 
         4 e^{2 t \gamma } t \gamma  )  {\sigma }^4 + 
      ( -1 + e^{2 t \gamma } )  {\gamma }^3 {\sigma }^6 ) }{2 {( D ( -1 + e^{2 t \gamma } )  + 
        \gamma  {\sigma }^2 ) }^4}$\\
$\beta_{1,4,0}(t)$&$=\mystyle\frac{-( e^{2 t \gamma } 
      ( D^3 ( 2 - 6 e^{4 t \gamma } + e^{6 t \gamma } + 
           3 e^{2 t \gamma } ( 1 + 4 t \gamma  )  )  - 6 D^2 \gamma  ( 1 - e^{4 t \gamma} + 
           4 e^{2 t \gamma } t \gamma  )  {\sigma }^2 + 
        6 D {\gamma }^2 ( 1 + 
           e^{2 t \gamma} ( -1 + 2 t \gamma )  )  
         {\sigma }^4 + 2 ( -1 + e^{2 t \gamma } )  
         {\gamma }^3 {\sigma }^6 )  ) }{4 
    {( D ( -1 + e^{2 t \gamma } )  + 
        \gamma  {\sigma }^2 ) }^4}$\\
\end{tabular}
\caption{\label{tab:coeff2}First order cumulant expansion coefficients $\beta_{1,k,l}(t)$.}
\end{table}
\end{center}}
\end{widetext}
They are plotted for $\gamma > 0$ 
in Figure \ref{Result} where we distinguish three time regimes from their
qualitative behavior. In the limit $t \to 0$ all expansion coefficients $\beta_{1,k,l}(t)$ vanish as already the 
probability density of the Brownian motion (\ref{harm-sol1}) in (\ref{kumulant-ansatz})--(\ref{coeff-chiansatz2}) 
leads to the correct initial distribution (\ref{harm-normal}). In the opposite limit $t \to \infty$ the only nonvanishing
expansion coefficients $\beta_{1,k,l}(t)$ read
\begin{eqnarray}
\label{NONV}
\hspace*{-0.2cm} \gamma > 0: \hspace*{0.2cm} \lim_{t \to \infty} \beta_{1,0,0}(t) = \frac{3 D}{4 \gamma^2} \, , \quad
\lim_{t \to \infty} \beta_{1,4,0}(t) = - \frac{1}{4D} \, , 
\end{eqnarray}
so that (\ref{kumulant-ansatz})--(\ref{coeff-chiansatz2}) reproduces the correct stationary solution (\ref{STAT}), (\ref{def:nweak}) up to first order in $g$. For intermediate times $t$ we observe that all expansion coefficients $\beta_{1,k,l}(t)$
show a nontrivial time dependence.
Note that the number of coefficients $\beta_{n,k,l}(t)$ which have to be calculated in the $n$th order is given by
$\sum_{k=0}^{n+1} (2k+1)=(n+2)^2$, thus it increases quadratically. The expansion coefficients $\beta_{n,k,l}(t)$
up the 7th order can be found in Ref. \cite{internet}.
\section{\label{sec:vpt}VARIATIONAL PERTURBATION THEORY}
In this section we follow Refs. \cite{Putz,Dreger} and perform a variational resummation of the cumulant expansion in close 
analogy to Section 
\ref{VPTNORM}. By doing so, we variationally calculate the probability density $\mP(x,t)$ for an arbitrarily large
coupling constant $g$ with $\gamma > 0$ (anharmonic oscillator) and $\gamma < 0$ (double well). In both cases, we obtain
probability densities, which originally peaked at the origin, turn into their 
respective stationary solutions in the long-time limit.
\subsection{Resummation Procedure}
We aim at approximating the nonlinear drift coefficient (\ref{ANHARM}) by the linear one $- \kappa x$ of a
trial Brownian motion with a damping coefficient $\kappa$ which we regard as our variational parameter. 
To this end we add $- \kappa x$ to the nonlinear drift
coefficient (\ref{ANHARM}) and subtract it again:
\begin{eqnarray}
\label{DECO}
K ( x ) = - \kappa x - g x^3 - \left( \gamma - \kappa \right) x \, .
\end{eqnarray}
By doing so, we consider the linear term $- \kappa x$ as the unperturbed system and treat all remaining terms in (\ref{DECO})
as a perturbation. Such a formal perturbation expansion is performed by introducing an artificial parameter $\delta$ which is
later on fixed by the condition $\delta=1$:
\begin{eqnarray}
\label{DECOB}
K ( x ) = - \kappa x - \delta \left[ g x^3+ \left( \gamma - \kappa \right) x \right] \, .
\end{eqnarray}
Performing the rescaling
\begin{eqnarray}
\label{RESC1}
x \to \frac{\tilde{x}}{\beta(\delta)}\, , \hspace*{0.6cm}
t \to \frac{\tilde{t}}{\beta^2(\delta)}\, , \hspace*{0.6cm}
g \to \frac{\beta^4(\delta)\tilde{g}}{\delta}\, 
\end{eqnarray}
with the scaling factor (\ref{BFAC}), the FP equation (\ref{FPE}) with the drift
coefficient (\ref{DECOB}) is transformed to the original one (\ref{FPNA}). Due to 
dimensional reasons also the parameter $\mu,\sigma$ of the initial distribution (\ref{coeff-p0t0}) have to be rescaled according to
\begin{eqnarray}
\label{RESC2}
\mu \to \frac{\tilde{\mu}}{\beta(\delta)}\, , \hspace*{0.6cm}
\sigma \to \frac{\tilde{\sigma}}{\beta(\delta)} \, .
\end{eqnarray}
The rescaling (\ref{RESC1}), (\ref{RESC2}) is applied to the cumulant expansion (\ref{kumulant-ansatz}), (\ref{p-def}).
After expanding in powers of $\delta$ and truncating at order $N$, we finally set $\delta=1$ and obtain some $N$th order
approximant $p^{(N)} ( x , t ;\kappa)$ for the cumulant.\\

Equivalently, the same result follows also from the weak-coupling expansion of the cumulant 
(\ref{kumulant-ansatz}), (\ref{p-def}):
\begin{eqnarray}
\label{WN}
p^{(N)} ( x , t ) = \ln \mP_{\gamma}(x,t) + \sum_{n=1}^N p_n ( x ,t ) g^n \, .
\end{eqnarray}
Treating in (\ref{DECO}) the linear drift coefficient $-\kappa x$ as the unperturbed term and all remaining terms
as a perturbation corresponds then to the substitution (\ref{SUB}) with the abbreviation (\ref{ABB}). Thus inserting
(\ref{SUB}) in (\ref{WN}), a reexpansion in powers of $g$ up to the order $N$ together with the resubstitution
(\ref{ABB}) leads also to  $p^{(N)} ( x , t ;\kappa)$ \cite{Putz}.\\

If we could have performed
this procedure up to infinite order, the variational parameter $\kappa$ would have dropped out of the expression, as the
original stochastic model (\ref{ANHARM}) does not depend on $\kappa$. However, as our calculation is limited to a finite
order $N$, we obtain an artificial dependence on $\kappa$, i.e.~some $N$th order approximant 
$p^{(N)} ( x , t ;\kappa)$, which has to
be minimized according to the principle of minimal sensitivity \cite{Stevenson}. Thus we search 
for local extrema of $p^{(N)} (x , t ; \kappa )$ with respect to $\kappa$, i.e.~from the condition
\begin{eqnarray}
\label{COND}
\left. \frac{\partial p^{(N)} (x , t ; \kappa )}{\partial \kappa}\right|_{\kappa=\kappa^{(N)}_{\rm opt} (x,t )} = 0 \, .
\end{eqnarray}
It may happen that this equation is not solvable within a certain region of the parameters $x, t$. In this
case we look for zeros of higher derivatives instead in accordance with the principle of minimal
sensitivity \cite{Stevenson}, i.e.~we determine the variational parameter $\kappa$ from solving
\begin{eqnarray}
\label{CONDB}
\left. \frac{\partial^m p^{(N)}(x , t ; \kappa )}{\partial \kappa^m}\right|_{\kappa=\kappa^{(N)}_{\rm opt,m} (x,t )} = 0 \, .
\end{eqnarray}
The solution $\kappa^{(N)}_{\rm opt} (x,t )$ from (\ref{COND}) or (\ref{CONDB}) yields the variational result
\begin{eqnarray}
\label{SOLV}
P ( x,t ) \approx 
\frac{\exp \left[ p^{(N)}\left(  x,t ; \kappa^{(N)}_{\rm opt} (x,t )\right)\right]}{\displaystyle \int_{-\infty}^{+ \infty} 
\exp \left[ p^{(N)}\left(  x',t ; \kappa^{(N)}_{\rm opt} (x',t )\right)\right] d x'}
\end{eqnarray}
for the probability density. Note that variational perturbation theory does not preserve the normalization of the
probability density. Although the perturbative result is still normalized in the
usual sense to the respective perturbative order in the coupling constant $g$, 
this normalization is spoilt by choosing an $x$-dependent damping constant $\kappa^{(N)}_{\rm opt} (x,t )$.
Thus we have to normalize the variational probability density according to (\ref{SOLV}) 
at the end \cite{Putz} (compare the similar situation for the variational ground-state wave function in Refs. \cite{Florian,Kunihiro}).\\
\subsection{Anharmonic Oscillator $(\gamma>0)$}
The variational procedure described in the last section is now applied for determining the time evolution of 
the probability density in case of the
nonlinear drift coefficient (\ref{ANHARM}) with $\gamma > 0$.
By doing so, the optimization
of the variational parameter $\kappa$ is performed for each value of the variable $x$ at each time $t$. 
The result of such a calculation 
with the parameters $g=D=1$, $\gamma=\sigma=0.1$ and $\mu=0$
is shown in Figure \ref{fig:results_par1_anharm} for the time interval $0 \leq t \leq 20$.
Figure \ref{fig:results_par1_anharm} (a) depicts the optimal variational parameter $\kappa$ which is the largest solution from (\ref{COND}). 
For small times, the optimal
variational parameter reveals a $x^2$-dependence. For large values of $t$, the variational parameter becomes
independent of $x$ which corresponds to the expectation that the probability density converges towards the stationary one.
In Figure \ref{fig:results_par1_anharm} (b) the respective variational results for the distribution are compared with the cumulant expansion.
The latter shows insofar a wrong behavior as it has two maxima whereas the numerical solution has only one. The variationally
determined probability density is depicted by dots which correspond to the optimal variational parameters in Figure \ref{fig:results_par1_anharm} (a).
We observe on this scale nearly no deviation between the variational and the numerical distribution, thus already the first-order
variational results are quite satisfactory. \\

\ignore{%
\begin{widetext}
\begin{center}
\begin{figure}[h!]
\includegraphics[scale=0.9]{figure6.eps} 
\caption{\label{fig:results_par1_anharm} Time evolution of probability density from variational optimization for $g=D=1$, $\gamma=\sigma=0.1$, 
and $\mu=0$. (a) Largest variational parameter $\kappa$ determined from (\ref{COND}), colors code different times corresponding to
the distributions shown in (b).
(b) Dots correspond to variational parameters of (a) and coincide on this scale
with numerical solutions of FP equation as represented by the lines through the dots. Cumulant expansions are shown as gray
areas. At the front the stationary distribution $\mPs(x)$ and the corresponding potential $\Phi(x)$ are depicted.} 
\end{figure}
\end{center}
\end{widetext}
}
\begin{widetext}
\begin{center}
\begin{figure}[h!]
\hbox{%
\mbox{\hbox to 0pt{\hspace{.1cm}(a)}%
\vtop{\vskip-2ex\hbox{\includegraphics[width=.32\linewidth]{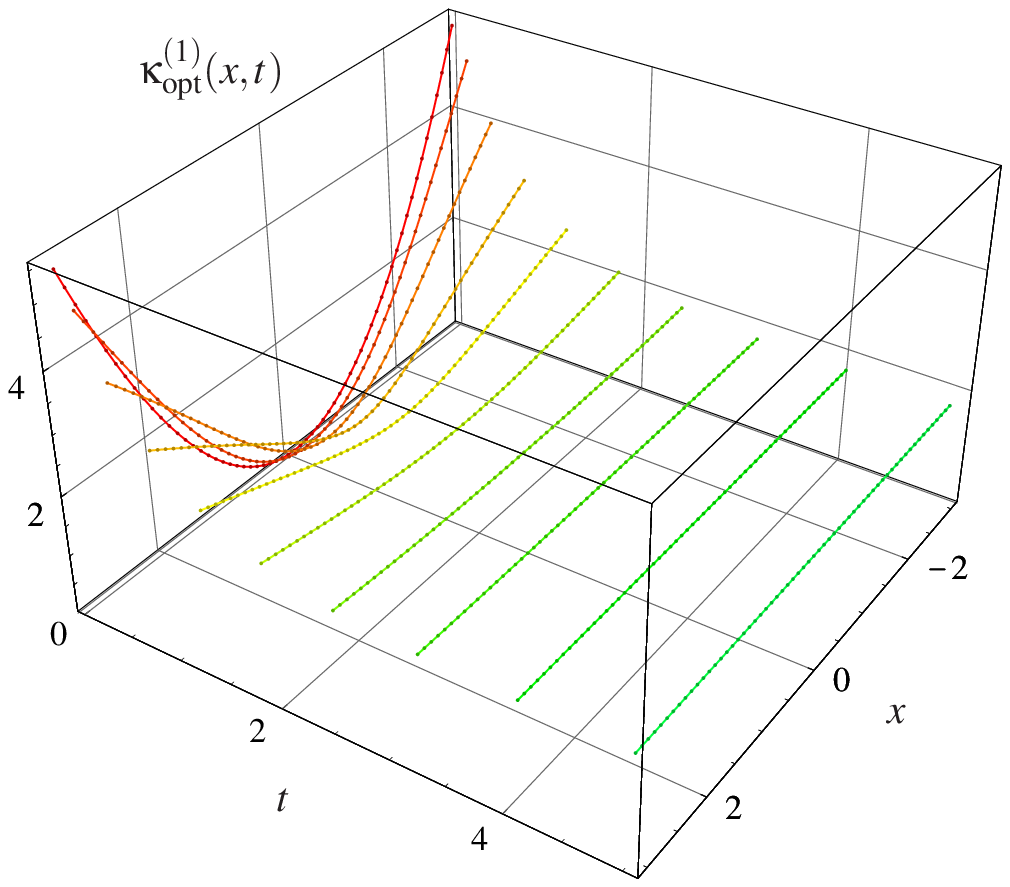}}}}
\mbox{\hbox to 0pt{\hspace{.1cm}(b)}%
\vtop{\vskip-2ex\hbox{\includegraphics[width=.32\linewidth]{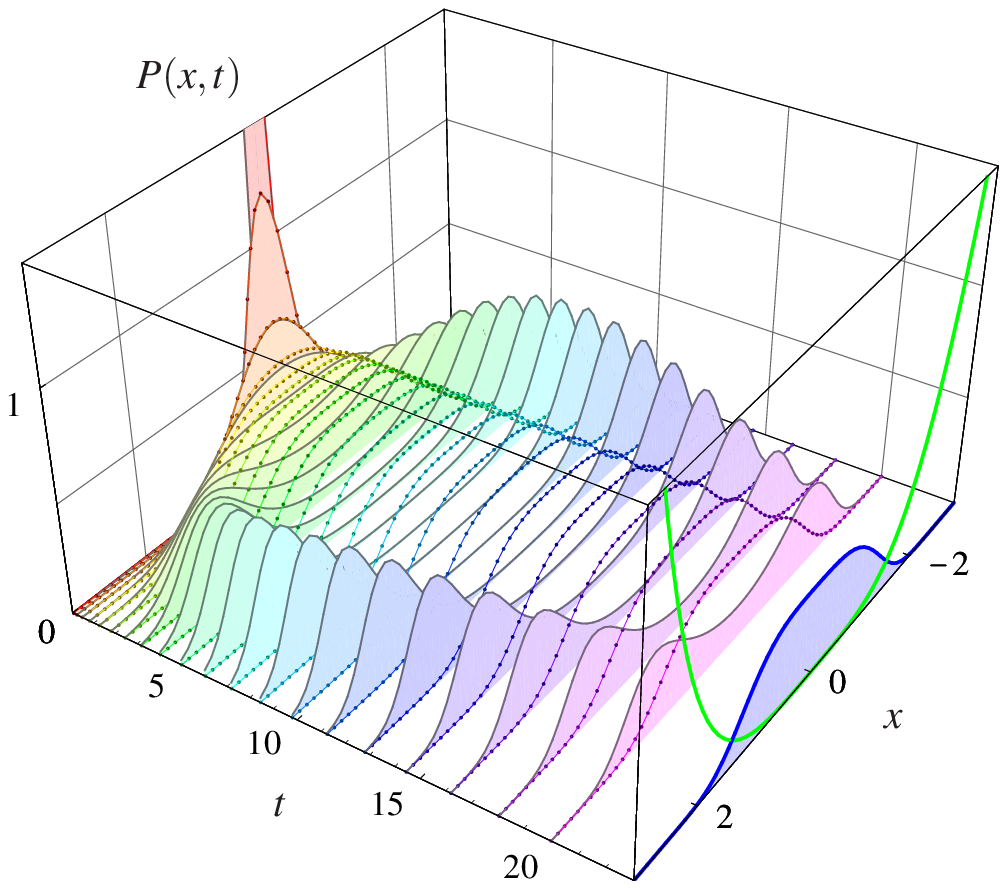}}}}
\mbox{\hbox to 0pt{\hspace{.0cm}(c)}%
\vtop{\vskip1ex\hbox{\includegraphics[width=.32\linewidth]{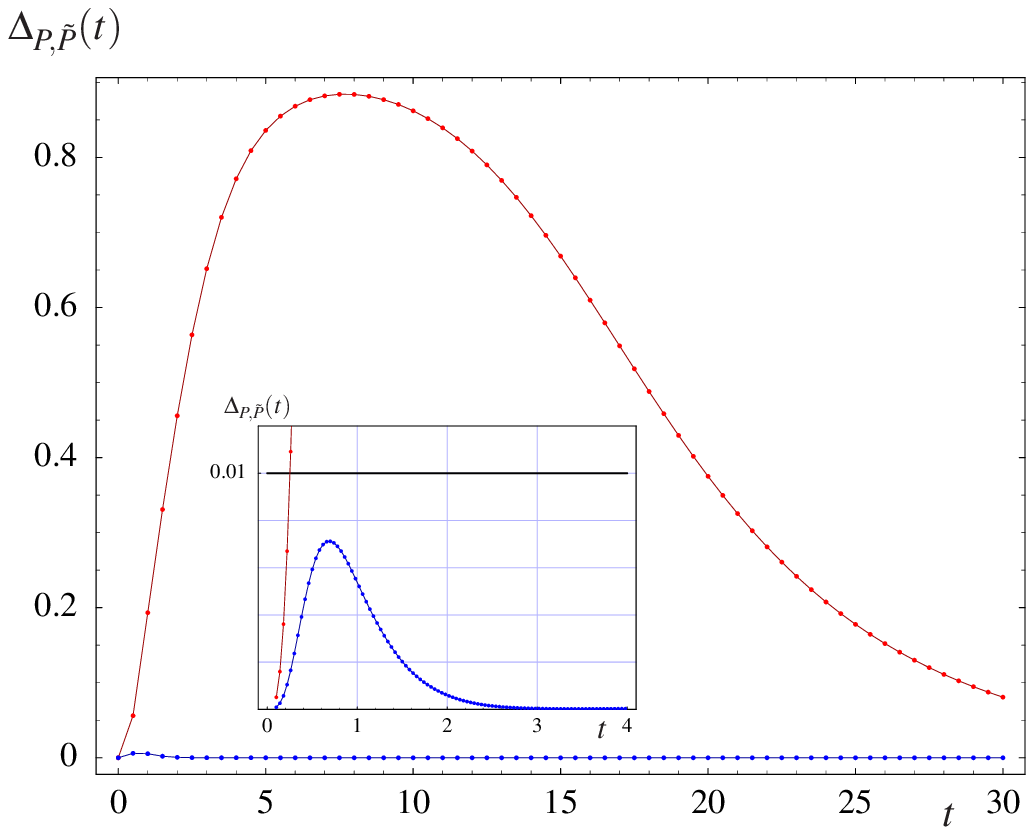}}}}
}%
\caption{\label{fig:results_par1_anharm} Time evolution of probability density from variational optimization for $g=D=1$, $\gamma=\sigma=0.1$, 
and $\mu=0$. (a) Largest variational parameter $\kappa$ determined from (\ref{COND}), colors code different times corresponding to
the distributions shown in (b).
(b) Dots correspond to variational parameters of (a) and coincide on this scale
with numerical solutions of FP equation as represented by the lines through the dots. Cumulant expansions are shown as gray
areas. At the front the stationary distribution $\mPs(x)$ and the corresponding potential $\Phi(x)$ are depicted. (c) Distance (\ref{dist}) between variational and cumulant expansion from numerical solution of FP equation.}
\end{figure}
\end{center}
\end{widetext}
In order to quantify the quality of our approximation, we introduce the distance 
between two distributions $\mP(x,t)$ and $\tilde{\mP}(x,t)$ at time $t$ according to
\begin{eqnarray}
\label{dist}
\Delta_{\mP,\tilde{\mP}} (t) = \frac{1}{2} \int_{-\infty}^{+\infty} \left| \mP(x,t) - \tilde{\mP}(x,t)\right| d x \, .
\end{eqnarray}
If both distributions are normalized and positive, the maximum value for the distance $\Delta_{\mP,\tilde{\mP}} (t)$ 
is $1$ and corresponds
to the case that the distributions have no overlap. However, if they coincide for all $x$, the distance 
$\Delta_{\mP,\tilde{\mP}} (t)$ vanishes.
Thus small values of $\Delta_{\mP,\tilde{\mP}} (t)$ indicate that both distributions are nearly identical. 
In Figure \ref{fig:results_par1_anharm} (c) we compare the time evolution of the distance (\ref{dist}) between the variational distribution
and the cumulant expansion from the numerical solution of the corresponding FP equation, respectively.  
We observe that the variational optimization leads for all times to better results than the cumulant expansion,
both being indistinguishable for very small and large times, as expected.
\subsection{Double Well $(\gamma<0)$}
\begin{figure}[t]
\jmdbox{\includegraphics[width=\columnwidth]{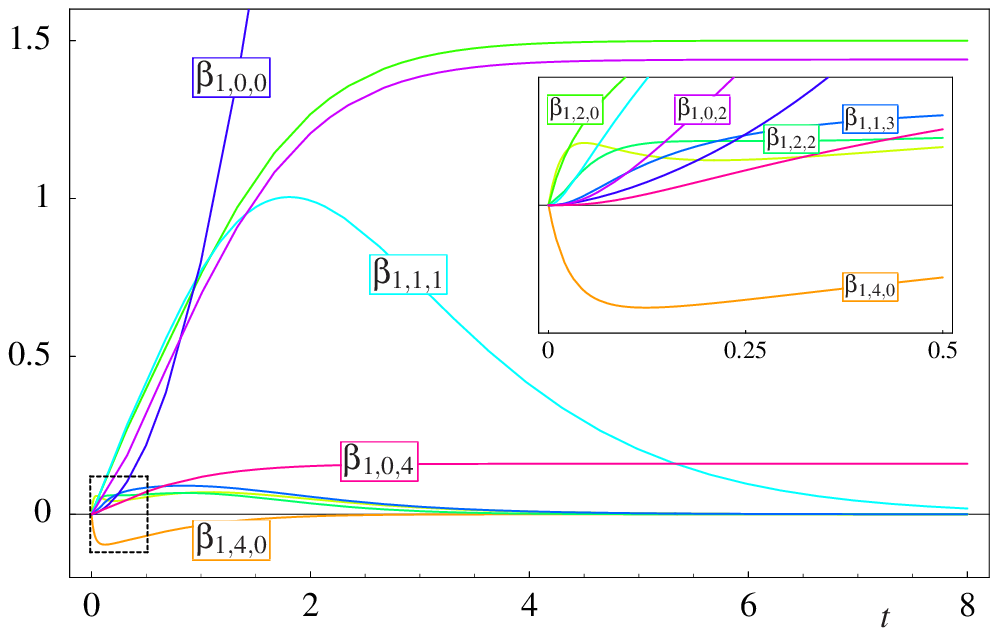}}%
\caption{\label{ResultB}Time evolution of the expansion coefficients $\beta_{1,k,l}(t)$ 
from Table II for $D=1$, $\gamma=-1$, $\mu=0$, 
and $\sigma=0.5$.}
\end{figure}

\begin{figure}[t]
\jmdbox{\includegraphics[width=.8\columnwidth]{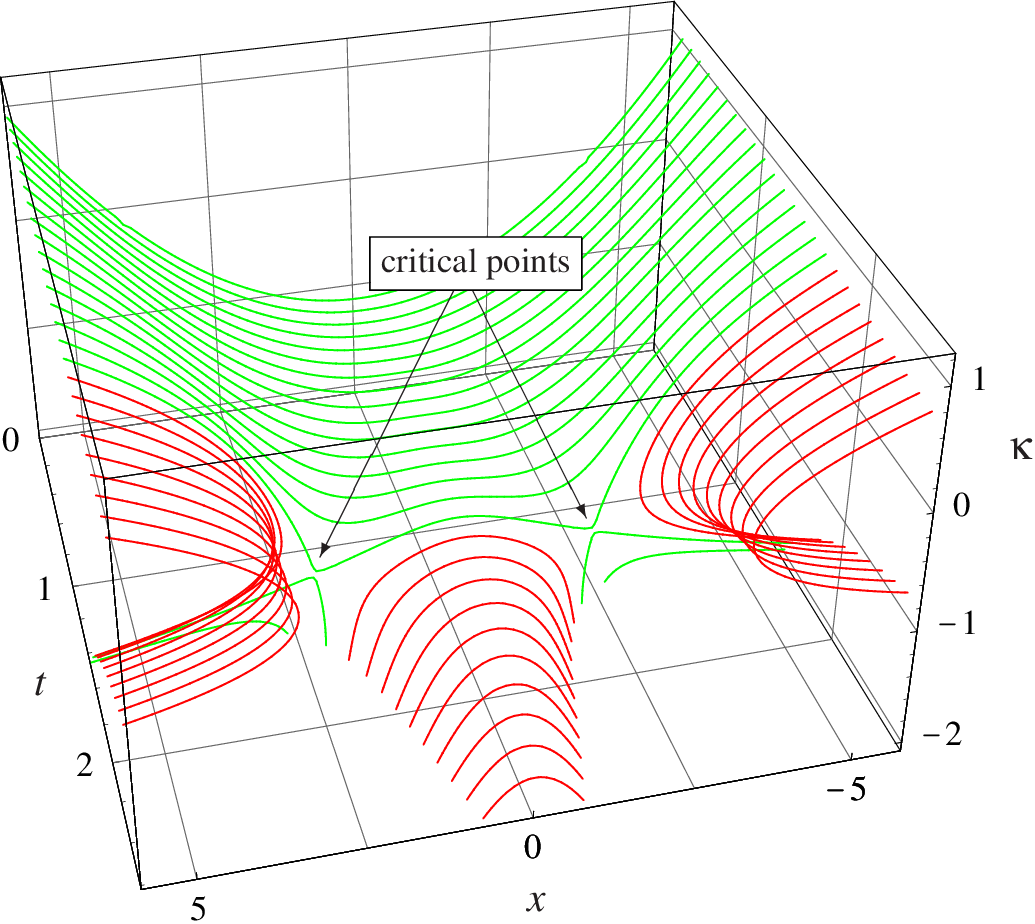}}%
\caption{\label{CritPlot}Zeros of $\partial p^{(1)} (x , t ; \kappa )/ \partial \kappa$ 
with continuous solutions for $t<t_{\rm crit}=1.618$ and
disconnected branches beyond.}
\end{figure}

\begin{figure}[h]
\jmdbox{\includegraphics[width=.8\columnwidth]{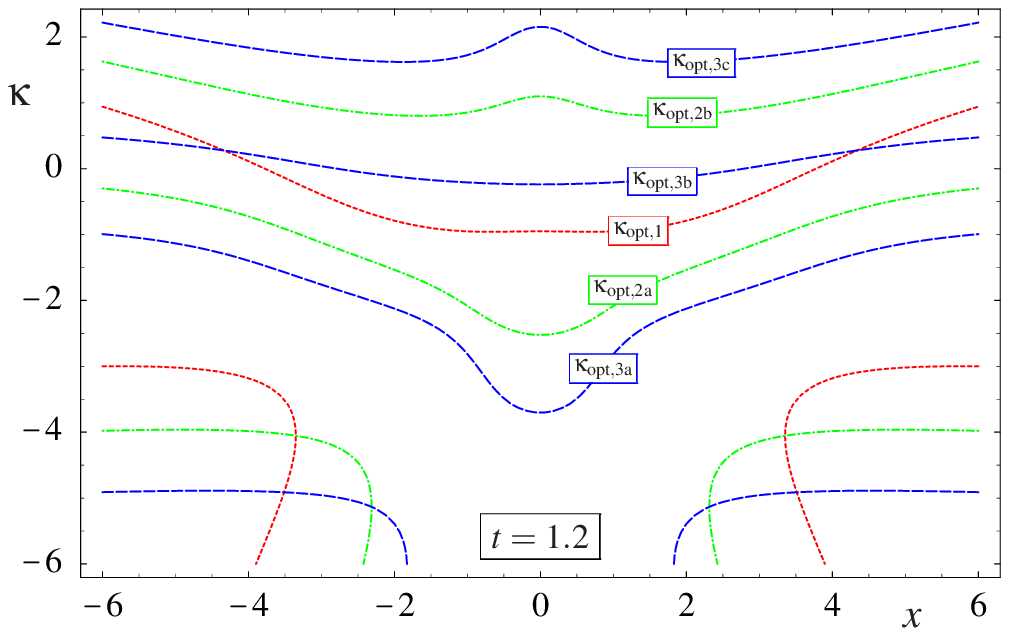}}%
\caption{\label{fig:zero3t12}Zeros of \eqref{CONDB} for $m=\{1,2,3\}$ at $t=1.2$, before reaching any critical time.}
\end{figure}

\begin{figure}[h]
\jmdbox{\includegraphics[width=.8\columnwidth]{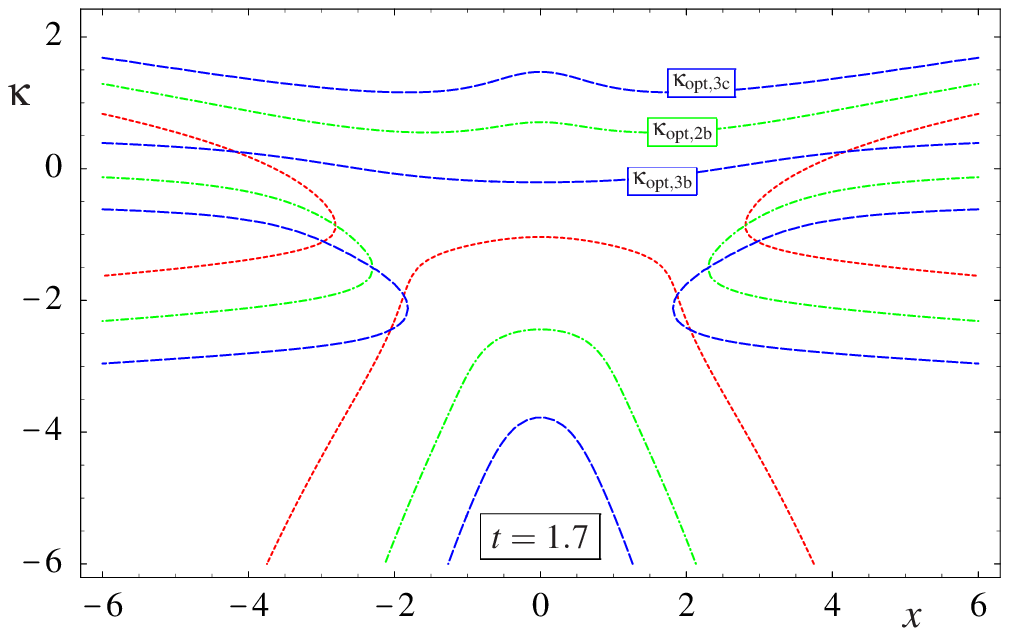}}%
\caption{\label{fig:zero3t17}Zeros of \eqref{CONDB} for $m=\{1,2,3\}$ at $t=1.7$. The smallest branch of zeros for
each derivative has reached a critical point and dissected in disconnected branches.}
\end{figure}

Now we discuss the more complicated problem of a nonlinear drift 
coefficient (\ref{ANHARM}) with $\gamma < 0$. The corresponding potential 
(\ref{NONPO}) has the form of a double well, i.e.~it decreases 
harmonically for small $x$ and becomes positive again for large $x$, 
so that a stationary solution exists (see the front of Figure \ref{fig:results_par2_double}(b)). 
Strictly speaking, the cumulant expansion developed in Section \ref{CE} 
makes no sense for $\gamma < 0$ as the unperturbed system $g=0$ does not have
a normalizable solution. This problem is reflected in the time evolution
of the cumulant expansion coefficients $\beta_{n,k,l}(t)$ which are listed in Table \ref{tab:coeff2}
and  depicted in Figure \ref{ResultB} for the order 
$n=1$. In contrast to the case $\gamma > 0$ in (\ref{NONV}), the coefficient
$\beta_{1,4,0}(t)$ vanishes for $\gamma < 0$ so that the cumulant expansion 
(\ref{kumulant-ansatz})--(\ref{coeff-chiansatz2}) does not even 
lead to the correct stationary solution (\ref{STAT}), (\ref{def:nweak}) 
up to first order in $g$. In the limit $t \to \infty$ the only nonvanishing 
expansion coefficients $\beta_{1,k,l}(t)$ read
\begin{eqnarray}
\gamma>0:
&& \lim_{t \to \infty}  \beta_{1,0,2}(t)= \frac{3D^2(D-2\gamma\sigma^2)}{2\gamma(D-\gamma\sigma^2)} \, , \hspace*{2mm}
\lim_{t \to \infty} \beta_{1,2,0}(t) = \frac{3}{2\gamma}\, , 
\nonumber \\&& 
\lim_{t \to \infty} \beta_{1,0,4}(t) =\frac{D^3}{4(D-\gamma \sigma^2)^4}\, .
\end{eqnarray}
\begin{widetext}
\begin{center}
\begin{figure}[h!]
\vskip0ex
\hbox{%
\mbox{\hbox to 0pt{\hspace{.1cm}(a)}
\vtop{\vskip-2ex\hbox{\includegraphics[width=.32\linewidth]{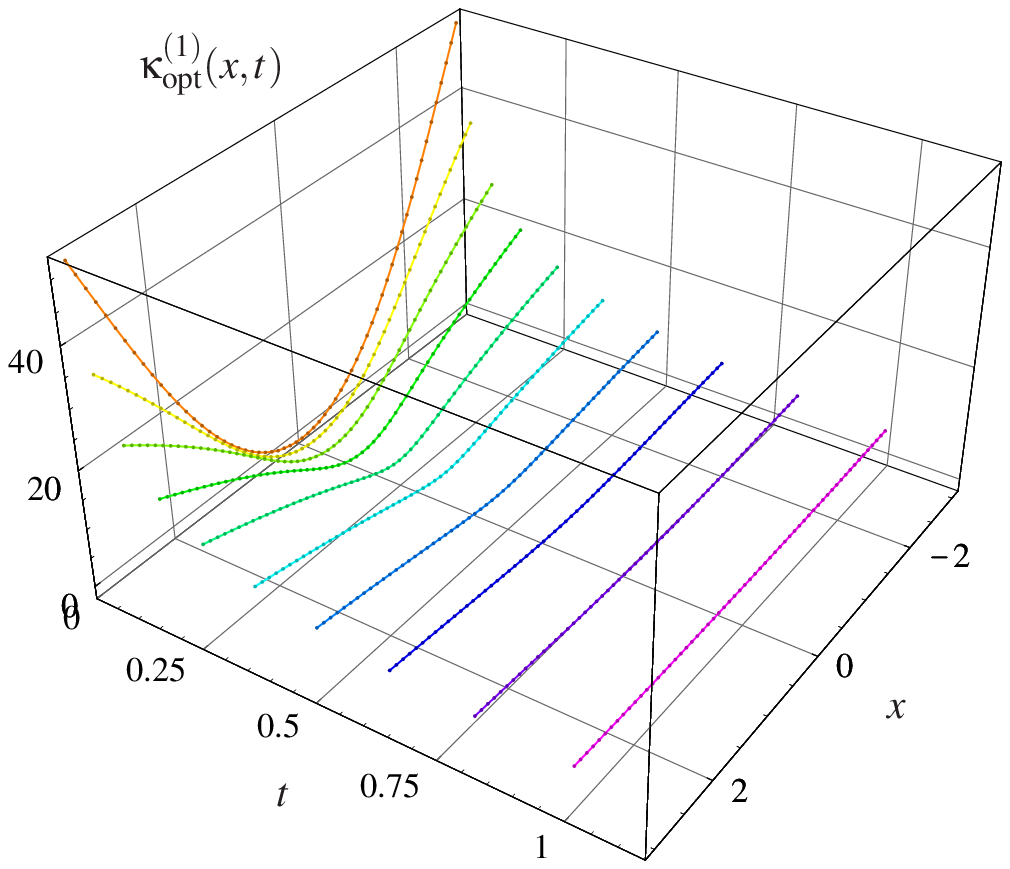}}}}
\mbox{\hbox to 0pt{\hspace{.1cm}(b)}
\vtop{\vskip-2ex\hbox{\includegraphics[width=.32\linewidth]{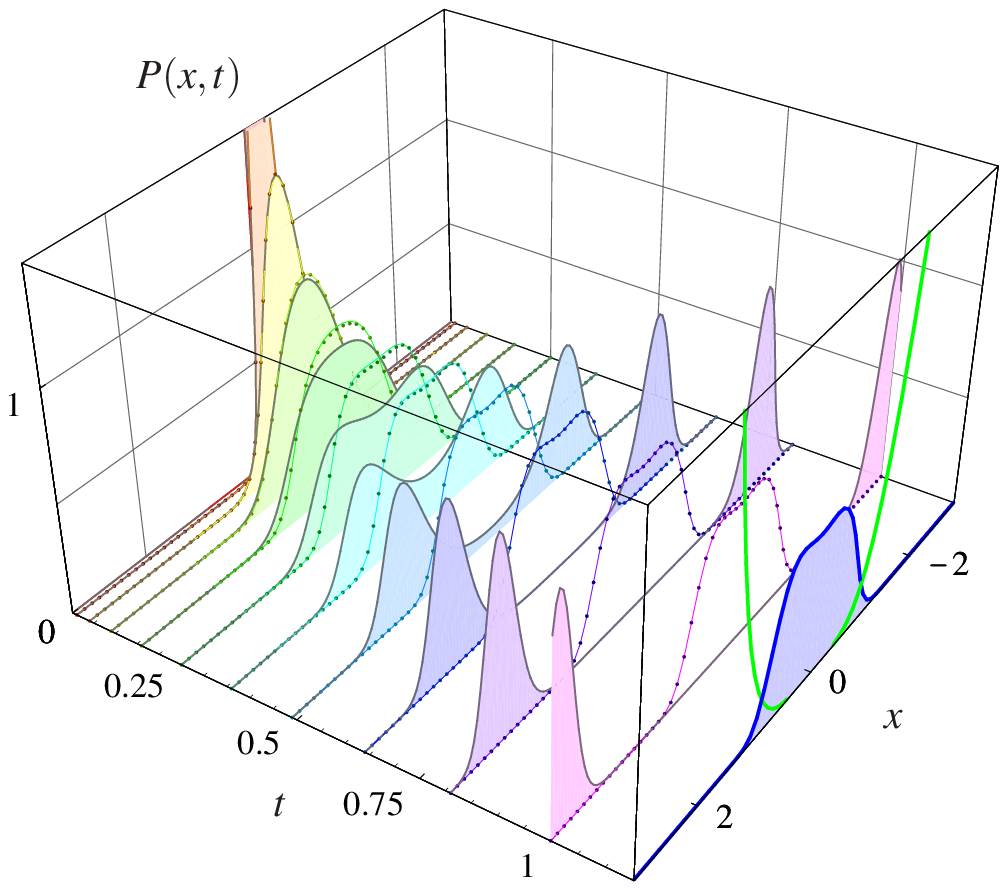}}}}
\mbox{\hbox to 0pt{\hspace{.0cm}(c)}
\vtop{\vskip1ex\hbox{\includegraphics[width=.32\linewidth]{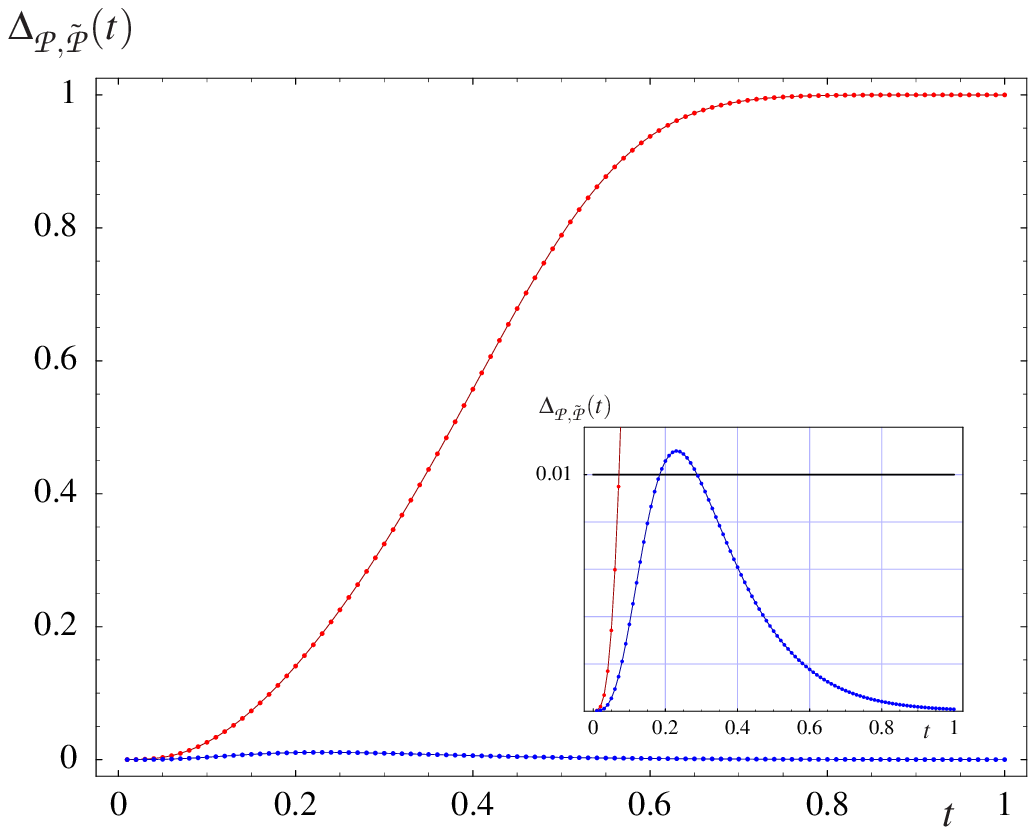}}}}
}%
\caption{\label{fig:results_par2_double} Time evolution of probability density from variational optimization for $g=10$, $D=1$, $\gamma=-1$,
$\sigma=0.1$ and $\mu=0$. 
(a) Largest variational parameter $\kappa$ determined from (\ref{COND}), colors code different times corresponding to
the distributions shown in (b).
(b) Dots correspond to variational parameters of (a) and coincide on this scale
with numerical solutions of FP equation as represented by the lines through the dots. 
Cumulant expansions are shown as gray areas. At the front the stationary distribution 
$\mPs(x)$ and the corresponding potential $\Phi(x)$ are depicted. 
(c) Distance (\ref{dist}) between variational and cumulant expansion from numerical solution of FP equation.}
\end{figure}
\end{center}
\end{widetext}
The results of the first-order variational calculation of the probability density are summarized in Figure \ref{fig:results_par2_double}
for the parameters $g=10$, $D=1$, $\gamma=-1$, $\sigma=0.1$,  and $\mu=0$. Due to the strong nonlinearity, we could determine 
the optimal variational parameter $\kappa$ from solving (\ref{COND}) for all $x$ and $t$ as shown in Figure \ref{fig:results_par2_double} (a).
As expected the cumulant expansion diverges for larger times $t$  as illustrated in Figure \ref{fig:results_par2_double} (b).
Despite of this the variational distribution lies precisely on top of the numerical solutions of the FP equation.
This impressive result is also documented in Figure \ref{fig:results_par2_double}(c) where the cumulant expansion shows for increasing time $t$
no overlap with the numerical solution, whereas the distance between the variational and the numerical distribution decreases.\\
Even more difficult is the case of a weak nonlinearity where the two minima of 
the double well are more pronounced. Therefore, the variational calculation
has also been performed for the parameter values $g=0.1$, $D=1$, $\gamma=-1$, $\sigma=0.1$, 
and $\mu=0$. For small times one obtains again a continuous optimal variational
parameter $\kappa^{(1)}_{\rm opt}(x,t)$ from solving (\ref{COND}) for all $x$. However, there exists
a critical time $t_{\rm crit} \approx 1.6$ beyond which condition (\ref{COND}) 
has different solution branches depending on $x$. 
In general such a critical point exists, if the equations:
\begin{equation}
\left[\begin{pmatrix}
1 \\
\partial/ \partial\kappa \\
\partial/ \partial x
\end{pmatrix}
\frac{\partial p^{(N)} (x , t ; \kappa )}{\partial \kappa}\right] _{\{x=x_{\rm crit}, t=t_{\rm crit}, \kappa=\kappa^{(N)}_{\rm crit}\}} =
\begin{pmatrix}
0 \\
0 \\
0
\end{pmatrix}
\end{equation}
have a solution $\{ x_{\rm crit}, t_{\rm crit}, \kappa_{\rm crit}\}$. For the case at hand we find
$x_{\rm crit} = \pm2.267$, $t_{\rm crit}=1.618$ and $\kappa_{\rm crit}=-1.326$. The corresponding surface of
zeros and the critical points are depicted in Figure~\ref{CritPlot}.
As it remains unclear how these branches should be combined for evaluating 
the probability density, we resort to zeros of higher derivatives \eqref{CONDB}
that are continuous for all times. For small values of $t$ we find one, two
and three continuous zeros for the first, second and third derivative, respectively,
as shown in Figure~\ref{fig:zero3t12} for $t=1.2$. The smallest zero
of the second and third derivative reach a critical point at $t_{\rm crit}=1.5284$ and
$t_{\rm crit}=1.4588$. The larger zeros have continuous solutions for
all values of $t$, as shown in Figure~\ref{fig:zero3t17} for $t=1.7$.\par
\begin{figure}[h]
\vskip10pt
\jmdbox{\includegraphics[width=.95\columnwidth]{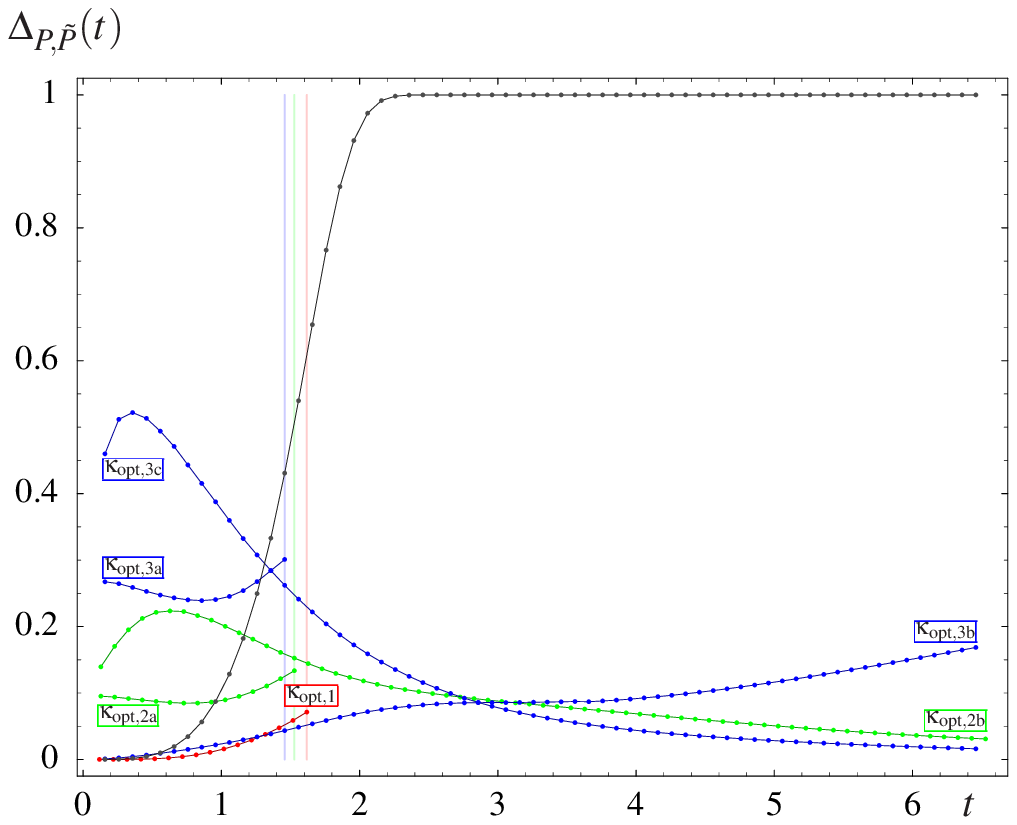}}%
\caption{\label{fig:plot3c}Distance \eqref{dist} between variational and cumulant 
expansion from numerical solution of FP equation for different branches of zeros. The smallest
zero of each derivative reaches a critical time at $t\approx1.6$.}
\end{figure}

Figure~\ref{fig:plot3c} shows the distance between variational and cumulant 
expansion from numerical solution of FP equation for these different branches of zeros.
Apparently there is no easy choice of the right branch of zeros, which gives
good results at all times. While $\kappa_{\rm opt,3b}$ gives good results at small times,
it fails to approach the stationary solution. The largest zero $\kappa_{\rm opt,3c}$ of the third derivative
on the other hand is unusable for small $t$, but gives good results for later times.\par
Assuming we had no knowledge of the numerical solution, we need to find
a way to switch the solution branch from $\kappa_{\rm opt,3b}$ to $\kappa_{\rm opt,3c}$. In order to
determine a suitable time to change the branch of zeros from $\kappa_{\rm opt}(x,t)$ to $\kappa'_{\rm opt}(x,t)$,
we consider the deviation of the moments of the corresponding distributions
\begin{eqnarray}
\label{momdist}
\Delta^n_{\kappa_{\rm opt},\kappa'_{\rm opt}}(t) = 
\frac{\int_{-\infty}^{+ \infty} \left| x^n \left[P(x,t;\kappa_{\rm opt}) - P(x,t;\kappa'_{\rm opt})\right]\right| d x}
{\int_{-\infty}^{+ \infty} \left| x^n P_{\rm stat}(x,t)\right| d x} \, .
\end{eqnarray}
\begin{figure}[h]%
\jmdbox{\includegraphics[width=.95\columnwidth]{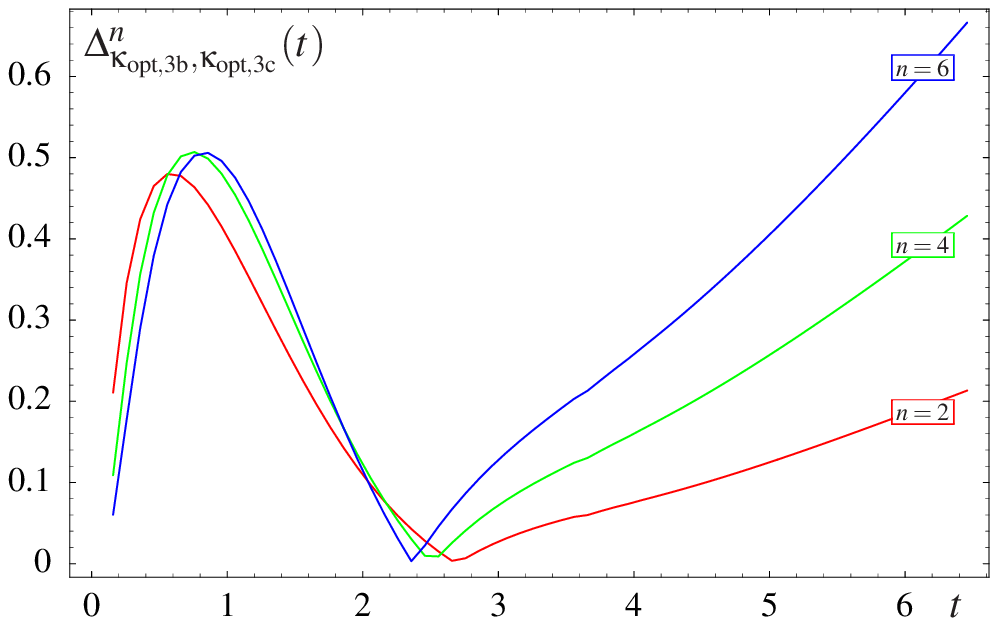}}%
\caption{\label{fig:plotmomdiff}Distance \eqref{momdist} between moments of distributions
determined from variationally from different branches of zeros of the third derivative.}
\end{figure}%
Figure \ref{fig:plotmomdiff} shows the distance \eqref{momdist} for the branches of zeros
$\kappa_{\rm opt,3b}(x,t)$ and $\kappa_{\rm opt,3c}(x,t)$ for the first three even moments $n={2,4,6}$.
We find, that the distributions are in good agreement at $t\approx2.6$, so we choose to
combine the solution for $\kappa_{\rm opt,3b}(x,t)$ for $t<2.6$ with the solution for
$\kappa_{\rm opt,3c}(x,t)$ for $t>2.6$. The combined result is shown in Figure \ref{fig:results_par3_double_hard}.
The distance (\ref{dist}) between variational expansion and numerical solution of the FP equation
for this case exhibits a small kink at $t\approx2.6$ due to the change in the
branch of zeros. Furthermore, in
comparison with the other cases in Figures \ref{fig:results_par1_anharm}
and \ref{fig:results_par2_double} this distance is relatively large which
underlines that this is, indeed, a difficult variational problem. Note, however,
that the combined solution succeeds in approaching the stationary solution
for large times.
\begin{widetext}
\begin{center}
\begin{figure}[h!]
\vskip1ex
\hbox{%
\mbox{\hbox to 0pt{\hspace{.1cm}(a)}
\vtop{\vskip-2ex\hbox{\includegraphics[width=.32\linewidth]{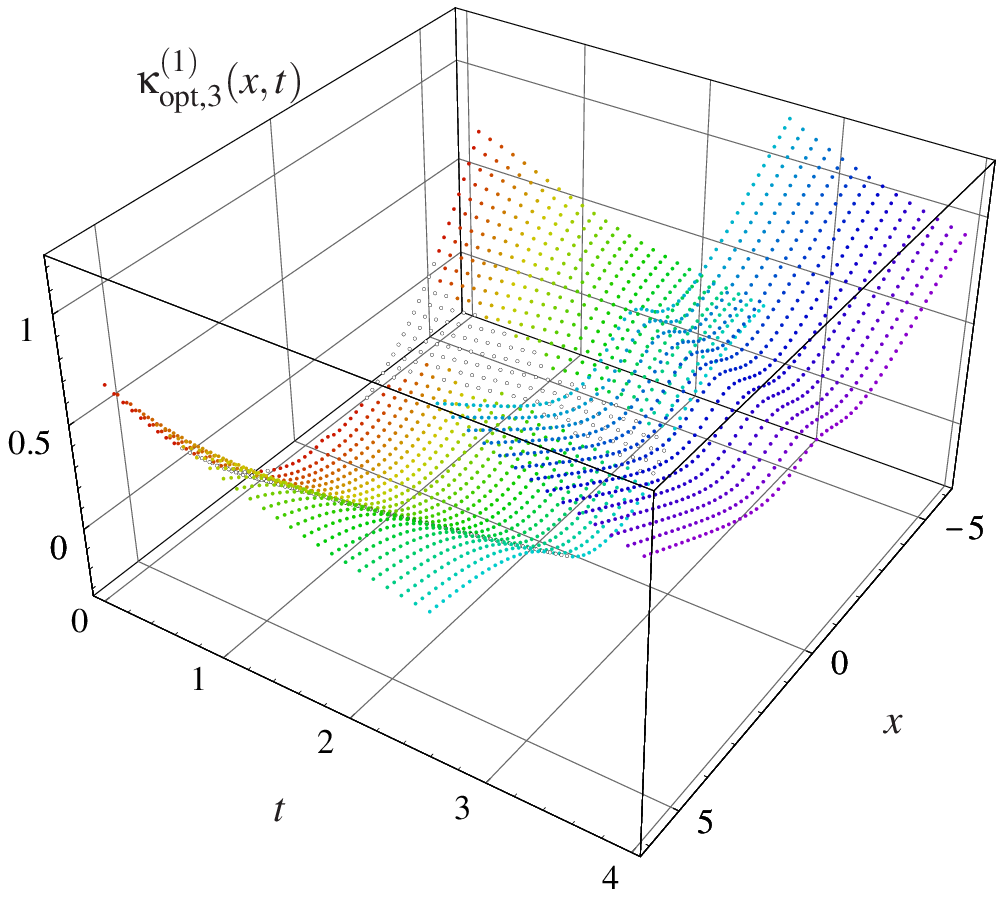}}}}
\mbox{\hbox to 0pt{\hspace{.1cm}(b)}
\vtop{\vskip-2ex\hbox{\includegraphics[width=.32\linewidth]{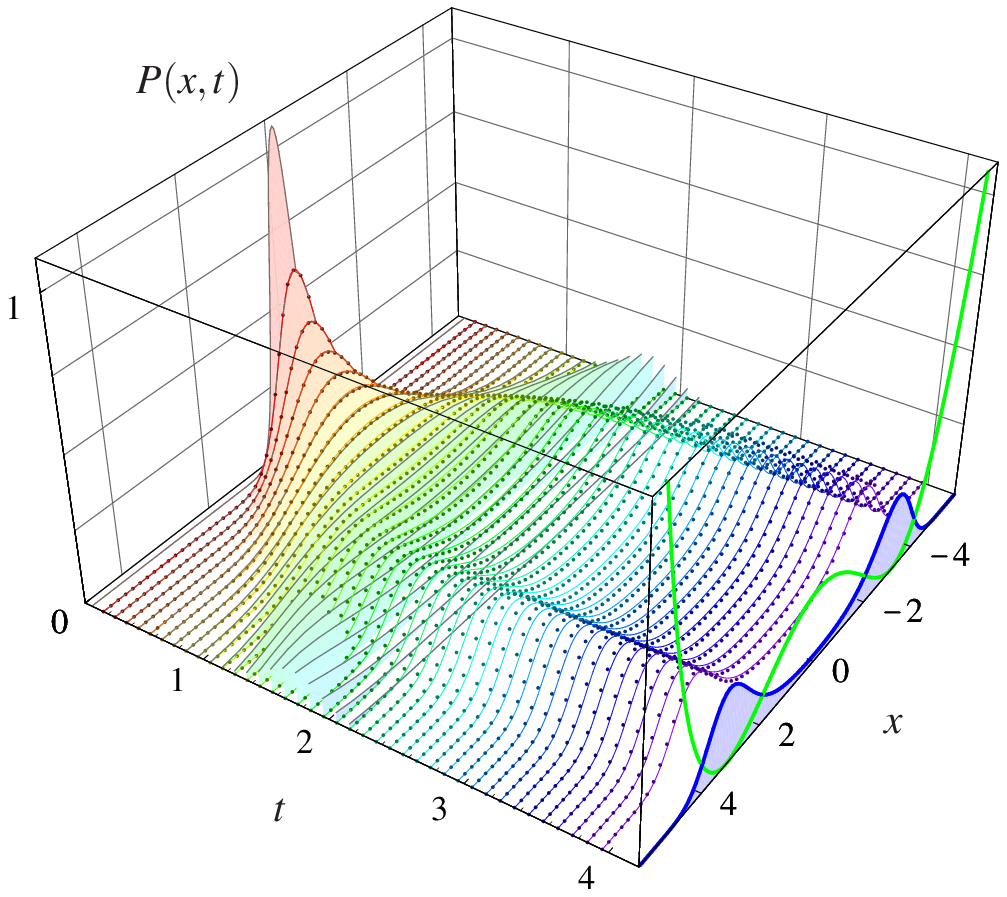}}}}
\mbox{\hbox to 0pt{\hspace{.0cm}(c)}
\vtop{\vskip1ex\hbox{\includegraphics[width=.32\linewidth]{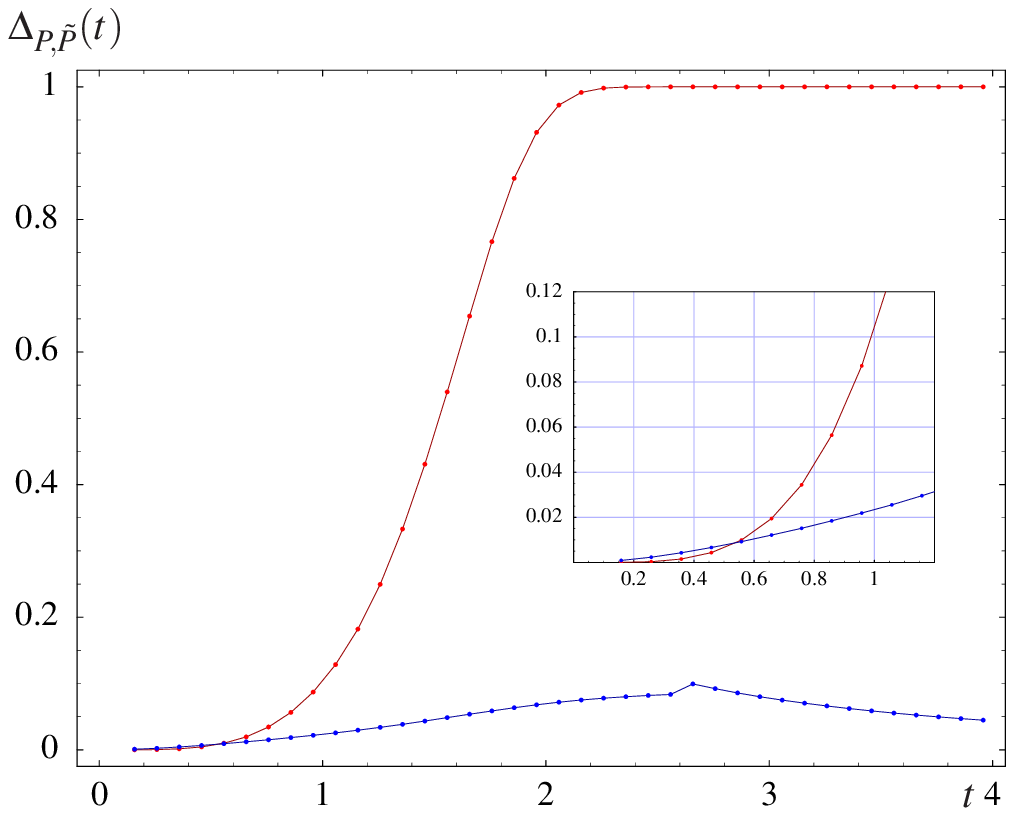}}}}
}%
\caption{\label{fig:results_par3_double_hard} Time evolution of probability density from variational optimization for 
$g=0.1$, $D=1$, $\gamma=-1$, $\sigma=0.1$,  and $\mu=0$.
(a) Variational parameter $\kappa$ determined from (\ref{CONDB}) for $m=3$, 
colors code different times corresponding to the distributions shown in (b).
For $t<2.6$ we selected the branch of zeros $\kappa_{\rm opt,3b}$, whereas for $t>2.6$
the branch $\kappa_{\rm opt,3c}$ was used. Outlined dots are interpolated zeros (see text).
(b) Dots correspond to variational parameters of (a) and coincide on this scale
with numerical solutions of FP equation as represented by the lines through the dots. 
Cumulant expansions are shown as gray areas. At the front the stationary 
distribution $\mPs(x)$ and the corresponding potential $\Phi(x)$ are depicted.
(c) Distance (\ref{dist}) between variational and cumulant expansion from numerical solution of FP equation.}
\end{figure}
\end{center}
\end{widetext}
We remark that the variational approach of Ref. \cite{Okopinska} is related to ours.
In contrast to our method, one obtains there in case of the difficult double
well problem $g=0.1$, 
$\kappa =-1$ 
a unique solution of the extremal condition (\ref{COND}) for all $x$ and $t$.
However, the resulting probability density shows for larger times $t$
significant deviations from our, and from numerical solutions
of the FP equation.
\subsection{Higher Orders}
\begin{center}
\begin{figure}[t!]
\vskip1ex
\jmdbox{\includegraphics[width=.95\columnwidth]{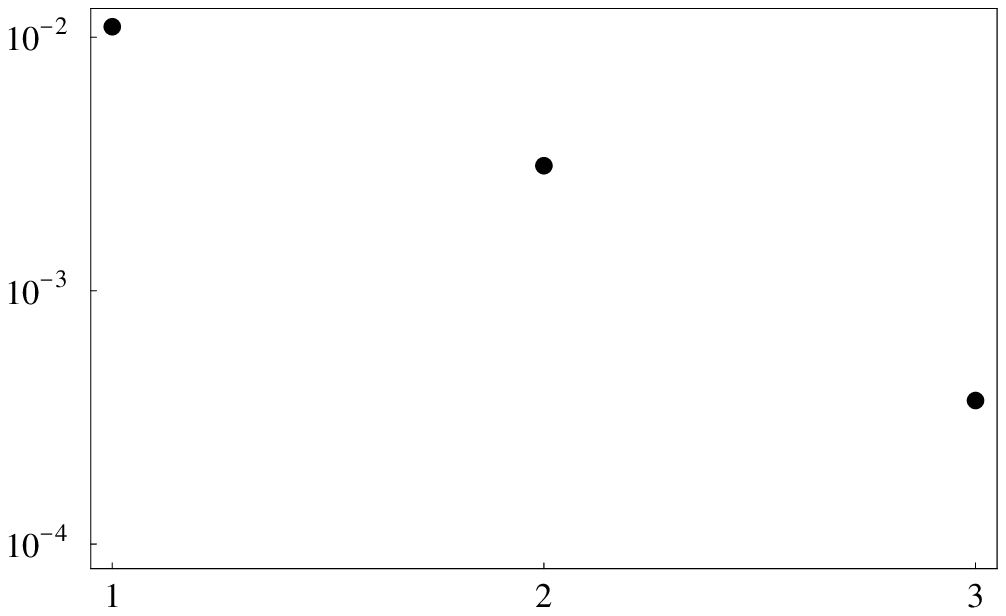}}%
\caption{\label{fig:hodouble} Distance (\ref{dist}) between numerical solution
of FP equation and the first three variational calculations
for $g=10$, $\gamma=-1$, $D=1$, $\sigma=0.1$, $\mu=0$ and $t=0.23$.} 
\end{figure}
\end{center}

High-order variational calculations have been performed for the double
well with the parameters $g=10$, $D=1$, $\gamma=-1$ in case of an initially
Gaussian-distributed probability density peaked at the origin, i.e.~$\sigma=0.1, \mu=0$
for $t=0.23$. This time was chosen due to its large distance between
the variational result and the numerical solution in order to reduce  
possible errors in the numerical solution. The order of magnitude of the
systematic error of the numerical solution can be estimated by comparing the numerical solution
of the harmonic problem, e.g. $g=0$, with the exact solution that is available
for that case. We find that the error of the numerical solution is about $10^{-6}$, which
is smaller that the pointsize used in Figure~\ref{fig:hodouble}.
The first three variational orders, shown in Figure \ref{fig:hodouble},
converge exponentially to the numerical solution of the FP equation.
\section{\label{sec:concl}SUMMARY}
We have presented high-order variational calculations for the probability density $\mP(x,t)$ of a stochastic model 
with additive noise which is
characterized by the nonlinear drift coefficient (\ref{ANHARM}). A comparison with numerical results shows an exponential
convergence of our variational resummation method with respect to the order. We hope that VPT will turn out to be useful
also for other applications in Markov theory as, for instance, the calculation of Kramer rates \cite{Haenggi1}
(see the recent variational calculation tunneling amplitudes from weak-coupling expansions in Ref. \cite{HampKleinert1}), 
the treatment of stochastic resonance \cite{Haenggi2}, or the investigation of Brownian motors \cite{Reimann}.
\begin{acknowledgments}
The authors thank Hagen Kleinert for fruitful discussions on variational perturbation theory.
\end{acknowledgments}


\begin{thebibliography}{99}
%
\bibitem{Gardiner}
C.W. Gardiner,
{\it Handbook of Stochastic Methods}, Second Edition 
(Springer, Berlin, 1985).
%
\bibitem{Stratonovich} 
R.L.~Stratonovich, 
{\it Topics in the Theory of Random Noise, Volume 1 -- General Theory of Random
Processes, Nonlinear Transformations of Signals and Noise,} Second Printing 
(Gordon and Breach, New York, 1967).
%
\bibitem{Kampen} 
N.G.~van~Kampen, {\it Stochastic Processes in
Physics and Chemistry} 
(North-Holland Publishing Company, New York, 1981).
%
\bibitem{Haken1} 
H.~Haken, 
{\it Synergetics -- An Introduction, Nonequilibrium Phase Transitions and Self-Organization
in Physics, Chemistry and Biology,} Third Revised and Enlarged Edition
(Springer, Berlin, 1983).
%
\bibitem{Risken}
H.~Risken, 
{\it The Fokker-Planck Equation -- Methods of Solution and Applications}, Second Edition 
(Springer, Berlin, 1988).
%
\bibitem{Haken3} 
H.~Haken, {\it Laser Theory}, Encyclopedia of Physics, Vol. XXV/2c
(Springer, Berlin, 1970).
%
\bibitem{HampKleinert1} 
B.~Hamprecht and H.~Kleinert, 
Phys. Lett. {\bf B 564}, 111 (2003).
%
\bibitem{Stevenson} 
P.M.~Stevenson, Phys. Rev. {\bf D 23}, 2916 (1981);
Phys. Rev. {\bf E 30}, 1712 (1985); 
Phys. Rev. {\bf E 32}, 1389 (1985);
P.M.~Stevenson and R.~Tarrach, 
Phys.~Lett. {\bf B 176}, 436 (1986).
%
\bibitem{Feynman1}
R.P. Feynman,
{\it Statistical Mechanics}
(Reading, Mas\-sa\-chu\-setts,1972).
%
\bibitem{Feynman2}
R.P.~Feynman and H.~Kleinert, 
Phys. Rev. {\bf A 34}, 5080 (1986).
%
\bibitem{Tognetti1}
R.~Giachetti and V.~Tognetti,
Phys. Rev. Lett. {\bf 55}, 912 (1985).
%
\bibitem{Tognetti2}
A. Cuccoli, R. Giachetti, V. Tognetti, R. Vaia, and P. Verrucchi,
J. Phys.: Condens. Matter {\bf 7}, 7891 (1995).
%
\bibitem{Kleinertsys}
H.~Kleinert,
Phys. Lett.  {\bf A 173}, 332 (1993).
%
\bibitem{Festschrift}
W. Janke, A. Pelster, H.-J. Schmidt, and M. Bachmann (Editors),
{\it Fluctuating Paths and Fields -- Dedicated to Hagen Kleinert
on the Occasion of His 60th Birthday} 
(World Scientific, Singapore, 2001). 
%
\bibitem{Kleinert} 
H.~Kleinert, 
{\it Path Integrals in Quantum Mechanics, Statistics, Polymer Physics, and Financial Markets},  Third Edition
(World Scientific, Singapore, 2004).
%
\bibitem{KleinertD1}
H. Kleinert,
Phys. Rev. {\bf 57}, 2264 (1998);
Addendum: Phys. Rev. D {\bf 58},  107702 (1998).
%
\bibitem{KleinertD2}
H. Kleinert,
Phys. Rev. {\bf D 60},  085001 (1999).
%
\bibitem{Verena}
H. Kleinert and V. Schulte-Frohlinde, 
{\it Critical Properties of $\phi^4$-Theories}
(World Scientific, Singapore, 2001).
%
\bibitem{Wegner}
F.J. Wegner, 
Phys. Rev. {\bf B 5}, 4529 (1972).
%
\bibitem{Satellit}
H. Kleinert,
Phys. Lett. {\bf A 277}, 205 (2000).
%
\bibitem{Lipa}
J.A.~Lipa, D.R.~Swanson, J.A.~Nissen, Z.K.~Geng, P.R.~Williamson, D.A.~Stricker, T.C.P.~Chui, U.E.~Israelsson, and M.~Larson, 
Phys. Rev. Lett. {\bf 84}, 4894 (2000). 
%
\bibitem{Putz} 
H.~Kleinert, A.~Pelster, and Mihai V.~Putz,
Phys.~Rev. {\bf E 65}, 066128 (2002).
%
\bibitem{Okopinska}
A. Okopinski\'{n}ska, 
Phys. Rev. {\bf E 65}, 062101 (2002).
%
\bibitem{Dreger}
J. Dreger, diploma thesis (in German), 
Free University of Berlin (2002).
%
\bibitem{Janke1}
W.~Janke and H.~Kleinert,
Phys. Rev. Lett. {\bf 75}, 2787 (1995).
%
\bibitem{Janke2}
H.~Kleinert and W.~Janke,
Phys. Lett. {\bf A 206}, 283 (1995).
%
\bibitem{Guida}
R.~Guida, K. Konishi, and H. Suzuki,
Ann. Phys. {\bf 249}, 109 (1996).
%
\bibitem{Proof}
H. Kleinert,
Phys. Rev. {\bf D 57}, 2264 (1998).
%
\bibitem{Kuerzinger}
H. Kleinert, W. K{\"u}rzinger, and A. Pelster, 
J. Phys. {\bf A 31}, 8307 (1998).
%
\bibitem{Weissbach} 
F.~Weissbach, A.~Pelster, and B.~Hamprecht, 
Phys.~Rev. {\bf E 66}, 036129 (2002).
%
\bibitem{Schanz}
A. Pelster, H. Kleinert, and M. Schanz,
Phys. Rev. {\bf E 67}, 016604 (2003).
%
\bibitem{Gradshteyn}
I.S.~Gradshteyn and I.M.~Ryzhik, {\it Table of Integrals, Series,
and Products,} Corrected and Enlarged Edition 
(Academic Press, New York, 1980).
%
\bibitem{Bender} 
C.~M.~Bender and T.~T.~Wu, 
Phys.~Rev. {\bf 184}, 1231 (1969); 
Phys.~Rev. {\bf D 7}, 1620 (1973).
%
\bibitem{internet}
The expansion coefficients $\beta_{n,k,l}(t)$ up to seventh order can be found at \\
{\tt http://www.physik.fu-berlin.de/{\~{}}dreger/coeffs/}.
%
\bibitem{Florian}
A.~Pelster and F.~Weissbach, {\it Variational Perturbation Theory for the Ground-State Wave Function},
in {\it Fluctuating Paths and Fields}, Eds. W. Janke, A. Pelster, H.-J. Schmidt, and M. Bachmann
(World Scientific, Singapore, 2001), p. 315; eprint: {\tt quant-ph/0105095}.
%
\bibitem{Kunihiro}
T. Hatsuda, T.~Kunihiro, and T. Tanaka,  
Phys. Rev. Lett. {\bf 78}, 3229 (1997).
%
\bibitem{Haenggi1}
P. H{\"a}nggi, P. Talkner, and M. Borkovec, 
Rev. Mod. Phys. {\bf 62}, 251 (1990).
%
\bibitem{Haenggi2}
L. Gammaitoni, P. H{\"a}nggi, P. Jung, and F. Marchesoni, 
Rev. Mod. Phys. {\bf 70}, 223 (1998).
%
\bibitem{Reimann}
P. Reimann,
Phys. Rep. {\bf 361}, 57 (2002).
%
\end{thebibliography}
\end{document}